\documentclass{aa}  

\usepackage{amsmath, amssymb, amsfonts, amscd}
\usepackage{natbib}
\usepackage{graphicx}
\usepackage{txfonts}
\usepackage[final]{hyperref}
\usepackage{mathrsfs}
\usepackage{color}
\usepackage{mathabx}
\usepackage{supertabular}
\usepackage{longtable}
\usepackage{ragged2e}
\usepackage{tikz}
\usepackage{multirow}
\usepackage{siunitx}
\usepackage{xspace}
\usepackage[switch]{lineno}

\hypersetup{
   colorlinks=true,
   citecolor=blue,
   linkcolor=blue,
   urlcolor=black
   }

\definecolor{emerald}{rgb}{0.0,0.5,0.0}
\definecolor{smcolor}{rgb}{0.7,0.3,0.0}
\definecolor{blue-violet}{rgb}{0.54, 0.17, 0.89}
\definecolor{royalblue}{rgb}{0.254,0.412,0.882}
\definecolor{mygreen}{rgb}{0,0.392,0}
\definecolor{orangered}{rgb}{1,0.271,0}
\definecolor{crimson}{rgb}{0.86,0.08,0.24}

\newcommand{\padc}[1]{}

\newcommand{\LEt}[1]{}

\def\inumber{i}                                 
\def\Zset{\mathbb{Z}}                   
\def\Rset{\mathbb{R}}                   
\def\sign{{\rm sign}}                           
\def\scale{\, \propto \,}                       

\def\cotan{{\rm cotan} \,}

\newcommand{\ddroit}{{\rm d}}                           
\newcommand{\vect}[1]{\boldsymbol{#1}}          
\newcommand{\evect}[1]{\vect{{\rm e}}_{#1}}     
\newcommand{\normvar}[1]{\tilde{#1}}  
\newcommand{\dd}[2]{\partial_{#2} #1}           
\newcommand{\ddd}[3]{\partial_{#2 #3} #1}       
\newcommand{\DD}[2]{\dfrac{\ddroit #1}{\ddroit #2}}                     
\newcommand{\DDn}[3]{\dfrac{\ddroit^{#3} #1}{\ddroit #2^{#3}}} 
\newcommand{\scal}[2]{\langle #1 , #2 \rangle}  
\newcommand{\dotprod}[1]{\cdot}
\newcommand{\abs}[1]{\left| #1 \right|}         

\newcommand{\infvar}[1]{\ddroit #1}

\newcommand{\kron}[2]{\delta_{#1, #2}}
\def\dotp{\cdot }

\def\ihoriz{{\rm h}}
\def\nab{\nabla}                                        
\def\grad{\nab}                                 
\def\gradh{\grad_{\ihoriz}}                     
\def\divh{\gradh \dotp}                 

\def\mm{m}                                      
\def\kk{k}                                              

\def\pp{p}
\def\Legendre{P}                                

\def\nn{n}                                              
\def\Hough{\Theta}                              
\def\spinpar{\nu}                        

\def\mnu{{\mm,\spinpar}}                        
\def\Laplace{\mathcal{L}^{\mnu}}        
\def\Lapcol{\mathcal{L}_{\col}^{\mnu}}
\def\Laplon{\mathcal{L}_{\lon}^{\mnu}}

\def\alphapar{\alpha}
\def\betapar{\beta}
\def\gammapar{\gamma}
\def\betasurf{\betapar_{\isurf}}
\def\Jsurf{J_{\isurf}}
\def\Usurf{\gravpot_{\iforcing ; \isurf}}

\newcommand{\LegFnorm}[2]{\Legendre_{#1}^{#2}}          
\newcommand{\expo}[1]{{\rm e}^{#1}}                             
\newcommand{\integ}[4]{\int_{#3}^{#4} #1 \infvar{#2} }   

\newcommand{\HoughF}[3]{\Hough_{#1}^{#2,#3}}            
\newcommand{\Houghcol}[3]{\Hough_{\col ; #1}^{#2,#3}}   
\newcommand{\Houghlon}[3]{\Hough_{\lon ; #1}^{#2,#3}}   
\newcommand{\Houghval}[3]{\Lambda_{#1}^{#2,#3}}         

\def\ipla{{\empty}}                              

\def\irot{\ipla}

\def\xp{q}
\def\xx{x}
\def\yy{y}
\def\zz{z}

\def\rr{r}                                           
\def\col{\theta}                                
\def\lon{\varphi}                             
\def\time{t}                                    

\def\er{\evect{\rr}}                            
\def\etheta{\evect{\col}}                       
\def\ephi{\evect{\lon}}                 

\def\ex{\evect{\xx}}
\def\ey{\evect{\yy}}
\def\ez{\evect{\zz}}

\def\framesymb{\mathcal{R}}                     

\newcommand{\framei}[1]{\framesymb_{#1}}

\def\framerot{\framei{\irot}}               

\newcommand{\rframe}[5]{\framesymb_{#1} {:} \left( #2, #3, #4, #5 \right) }

\def\spinrate{\Omega}                   
\def\spinvect{\vect{\spinrate}}         

\def\Rbody{R}                                   

\def\pressure{p}
\def\press{\pressure}
\def\Cp{C_\pressure}
\def\Cv{C_v}

\def\Rpla{\Rbody_\ipla}                 

\def\density{\rho}                              
\def\freq{\sigma}                               
\def\ggravi{g}                                  
\def\fnewton{\freq_{\rm C}}

\def\force{F}
\def\forcev{\vect{\force}}

\def\BVfreq{N_{\rm B}}

\def\csound{c_{\rm s}}
\def\temperature{T}
\def\calcvar{G}
\def\adiabexp{\Gamma_1}

\def\pressbg{\pressure_0}
\def\rhobg{\density_0}
\def\tempbg{\temperature_0}
\def\psurf{\pressure_{\rm s}}
\def\Tsurf{\temperature_\isurf}

\def\temp{\temperature}
\def\Tbulk{\overline{\temperature}}

\def\height{H}

\def\Rgp{R_{\rm IG}}
\def\Mmol{M_{\rm gas}}
\def\Rspec{R_{\rm s}}
\def\rcp{\kappa}

\def\heightsurf{\height_{\isurf}}

\def\ftide{\freq}                                       

\def\gravpot{U}                                 
\def\inputpow{J}

\def\gravpotn{\normvar{\gravpot}}
\def\vel{V}                                             
\def\isurf{{\rm s}}                             

\def\iforcing{{\rm T}}                          

\def\Utide{\gravpot_\iforcing}          

\def\Vvect{\vect{\vel}}                 
\def\Vr{\vel_\rr}                                       
\def\Vtheta{\vel_\col}                          
\def\Vphi{\vel_\lon}                            
\def\flux{F}

\newcommand{\vartide}[1]{\delta  #1}
\def\rhotide{\vartide{\density}}
\def\presstide{\vartide{\pressure}}
\def\temptide{\vartide{\temperature}}
\def\Gtide{\calcvar}
\def\Jtide{J}

\def\Vvectn{\normvar{\Vvect}}

\def\timen{\normvar{\time}}

\def\gradhn{\normvar{\grad}_{\ihoriz}}
\def\divhn{\gradhn \dotp}

\def\presstiden{\normvar{\presstide}}
\def\rhotiden{\normvar{\rhotide}}
\def\temptiden{\normvar{\temptide}}
\def\Gtiden{\normvar{\Gtide}}
\def\Jtiden{\normvar{\Jtide}}
\def\Vthetan{\normvar{\Vtheta}}
\def\Vphin{\normvar{\Vphi}}
\def\Vrn{\normvar{\Vr}}

\def\iref{0}

\def\speedref{\vel_{\iref}}

\def\phiwave{\phi}
\newcommand{\vprofilei}[3]{#1_{#2}}
\newcommand{\vprofile}[1]{\vprofilei{#1}{\nn}{\mm,\ftide}}
\def\presstidev{\vprofile{\presstiden}}

\def\rhotidev{\vprofile{\rhotiden}}
\def\temptidev{\vprofile{\temptiden}}
\def\Gtidev{\vprofile{\Gtiden}}
\def\Jtidev{\vprofile{\Jtiden}}
\def\Utidev{\gravpotn_{\iforcing; \nn}}
\def\Vthetav{\vprofile{\Vthetan}}
\def\Vphiv{\vprofile{\Vphin}}
\def\Vrv{\vprofile{\Vrn}}
\def\psiv{\vprofile{\Psi}}
\def\deptheq{\vprofile{h}}

\def\kwave{k}                           
\newcommand{\kverti}[1]{\kwave_{#1}}
\def\kvert{\kverti{\nn}}

\def\Utiden{\gravpotn_{\iforcing}}

\def\freson{\ftide_{{\rm L;} \nn}}

\def\Sn{S_{\nn}}

\def\Aconst{\mathcal{A}_\nn}        
\def\Bconst{\mathcal{B}_\nn}

\def\eigenval{\lambda}

\newcommand{\tidal}[1]{\delta #1}

\def\buoyancy{b}

\def\itherm{{\rm H}}

\def\bj{b_{\itherm}}

\def\dlnbet{\zeta}
\def\lambdapar{\lambda}

\def\psitherm{\Psi_{\nn;\itherm}}
\def\psigrav{\Psi_{\nn;{\rm G}}}
\def\fluxtiden{\tidal{\flux}_{\nn}}

\def\ekin{E_{\rm k}}
\def\qint{q}

\def\Lambn{\Houghval{\nn}{\mm}{\spinpar}}

\def\opacity{\eta}

\newcommand{\eq}[1]{Eq.~(\ref{#1})}
\newcommand{\eqs}[2]{Eqs.~(\ref{#1}) and~(\ref{#2})}
\newcommand{\eqsto}[2]{Eqs.~(\ref{#1}-\ref{#2})}
\newcommand{\eqsthree}[3]{Eqs.~(\ref{#1}), (\ref{#2}), and~(\ref{#3})}
\newcommand{\append}[1]{Appendix~\ref{#1}}

\newcommand{\fig}[1]{Fig.~\ref{#1}}

\newcommand{\sect}[1]{Sect.~\ref{#1}}

\newcommand{\comments}[1]{}

\newcommand{\papertwo}{Paper~II\xspace}
\newcommand{\paperthree}{Paper~III\xspace}

\usepackage{etoolbox}
\usepackage[refpage]{nomencl}

\providetoggle{nomsort}
\settoggle{nomsort}{true} 

\makeatletter
\iftoggle{nomsort}{%
    \let\old@@@nomenclature=\@@@nomenclature        
        \newcounter{@nomcount} \setcounter{@nomcount}{0}%
        \newcommand{\threedigits}[1]{\ifnum#1<100 0\two@digits{#1} \else \number#1\fi}
        \renewcommand\the@nomcount{\threedigits{\value{@nomcount}}}
        \def\@@@nomenclature[#1]#2#3{
          \addtocounter{@nomcount}{1}%
        \def\@tempa{#2}\def\@tempb{#3}%
          \protected@write\@nomenclaturefile{}%
          {\string\nomenclatureentry{\the@nomcount\nom@verb\@tempa @[{\nom@verb\@tempa}]%
          \begingroup\nom@verb\@tempb\protect\nomeqref{\theequation}%
          |nompageref}{\thepage}}%
          \endgroup
          \@esphack}%
      }{}
\makeatother

\newcommand{\mynomone}[3][section]{%
  \begingroup\edef\x{\endgroup
  \unexpanded{\nomenclature{#2}}%
    {\unexpanded{#3} \hspace*{\fill}  (\csname the#1\endcsname)}}\x}

\newcommand{\mynomtwo}[4][section]{%
  \begingroup\edef\x{\endgroup
  \unexpanded{\nomenclature[#2]{#3}}%
    {\unexpanded{#4} \hspace*{\fill}  (\csname the#1\endcsname)}}\x}

\renewcommand\nomgroup[1]{%
  \item[\bfseries
  \ifstrequal{#1}{A}{Acronyms}{%
  \ifstrequal{#1}{S}{Symbols}{%
  \ifstrequal{#1}{C}{Other Symbols}{}}}%
]}

\newcounter{logglabel}

\newcommand{\mynom}[3][S]{\nomenclature[#1]{#2}{#3~}}

\makenomenclature

\begin{document} 

\title{An analytical framework for atmospheric tides on rocky planets}
\subtitle{I. Formulation}

  \author{Pierre Auclair-Desrotour\inst{1} 
  \and Mohammad Farhat\inst{2,3,4} 
  \and Gwenaël Boué\inst{1} 
  \and Jacques Laskar\inst{1} 
          }

  \institute{LTE, Observatoire de Paris, Université PSL, Sorbonne Université, Univ. Lille, Laboratoire National de Métrologie et d'Essai, CNRS, 75014 Paris, France \\
    \email{pierre.auclair-desrotour@obspm.fr} 
    \and Department of Astronomy, University of California, Berkeley, Berkeley, CA 94720-3411, USA
    \and Department of Earth and Planetary Science, University of California, Berkeley, Berkeley, CA 94720-4767, USA
    \and Miller Fellow
  }

  \date{Received ...; accepted ...}

  \abstract
  {Atmospheric thermal tides arise from the diurnal contrast in stellar irradiation. They exert a significant influence on the long-term rotational evolution of rocky planets because they can accelerate the planetary spin, thereby counteracting the decelerating effect of classical gravitational tides. Consequently, equilibrium tide-locked states may emerge, as exemplified by Venus and hypothesised for Precambrian Earth.}
  {Quantifying the atmospheric thermal torque and elucidating its dependence on tidal frequency -- both in the low- and high-frequency regimes -- is therefore essential. In particular, we focus here on the resonance that affected early Earth, which is associated with a forced Lamb wave.}
  {Within the framework of linear theory, we develop a new analytical model of the atmospheric response to both gravitational an thermal tidal forcings for two representative vertical temperature profiles that bracket the atmospheres of rocky planets: (i) an isothermal profile (uniform temperature) and (ii) an isentropic profile (uniform potential temperature). Dissipative processes are incorporated via Newtonian cooling.}
  {We demonstrate that the isothermal and isentropic cases are governed by the same general closed-form solution, and we derive explicit expressions for the three-dimensional tidal fields (pressure, temperature, density and wind velocities) throughout the spherical atmospheric shell. These results constitute the foundation for two forthcoming papers, in which analytical formulae for the thermotidal torque will be presented and compared with numerical solutions obtained from General Circulation Models (GCMs).}
   {}


\maketitle



\section{Introduction}


Over the past two decades, the discovery of thousands of extrasolar planets has revealed the remarkable diversity of these worlds in terms of size, mass, and thermal state \citep[e.g.][]{Perryman2018book}. The launch of the {\it James Webb Space Telescope} \citep[JWST; e.g.][]{Gardner2006} in 2021 has ushered in a new era, offering unprecedented observational constraints on their atmospheric structure, climate, and composition -- particularly for temperate rocky planets orbiting within the habitable zones of low-mass stars \citep[e.g.][]{Kopparapu2013,Kopparapu2017,WK2022}. Because these characteristics are intrinsically linked to the long-term evolution of planetary systems, this evolution has become a subject of major scientific interest. The long-term orbital and rotational histories of exoplanets are largely governed by their tidal interactions with their host stars and moons \citep[e.g.][]{Hut1981,Rasio1996,Jackson2008}. It is therefore essential to understand how tides, including atmospheric thermal tides, influence the dynamical evolution of planetary systems.



Atmospheric thermal tides are global-scale waves arising from day-night differences in stellar radiation. Planetary atmospheres are heated on the dayside through the absorption of the incident stellar flux, and they cool down on the nightside, leading to periodic oscillations of pressure, temperature and wind velocities. These oscillations divide into two primary categories: migrating thermal tides and non-migrating thermal tides \citep[e.g.][]{CL70}. Migrating thermal tides designate waves following the sub-stellar point in its longitudinal motion with respect to the planet's surface, while non-migrating tides do not exhibit any global movement. Migrating thermal tides are of primary importance in celestial mechanics because they result in non-zero average torques influencing the planet's rotational evolution over million or billion years, analogous to gravitational tidal torques. The main contributor to this evolution is the semidiurnal thermal tide, namely the component associated with two oscillations within a stellar day.

Remarkably, semidiurnal thermotidal forces act to spin up the atmosphere in specific frequency intervals, thus opposing gravitational forces. Such a property results from the ability of gases to convert stellar radiation into mechanical energy, while gravitational tides only convert mechanical energy into heat. This allows thermal tides to drive planet-star systems away from their classical states of equilibrium, especially synchronisation, whereby the planet's rotation and orbital periods converge \citep[e.g.][]{Hut1980,Hut1981,Correia2008,Barnes2017}. As a result, thermal tides may generate differential rotation in the envelope of gaseous giants, such as hot Jupiters \citep[][]{AS2010,ADL2018,Gu2019,Lee2020}. However, they are mainly known for their key role in the long-term rotational evolution of rocky planets. In the latter, solid regions can support the loading caused by tidally-induced surface pressure fluctuations without being deformed, which precludes atmospheric mass excesses to be compensated. Consequently, thermal tides lead to non-negligible planetary-scale mass redistributions and torques on these planets. 

The thermodynamic acceleration of the Earth's rotation has been highlighted in the late 19th by Lord Kelvin, who found that it currently compensates ${\sim}10\%$ of the deceleration caused by semidiurnal gravitational tides \citep[][]{Thomson1882}. Modern estimates of the corresponding thermotidal torque derived from barometric measurements and simulations using general circulation models (GCMs)\mynom[A]{GCM}{general circulation model} range between $\SI{2.5e15}{J}$ and \SI{4.1e15}{J} \citep[e.g.][]{MM1975book,ZW1987,Schindelegger2014,Wu2023,DG2024}. The atmospheric torque is smaller than that induced by ocean tides, which generate $90\%$ of the tidally dissipated energy for the Lunar semidiurnal component \citep[e.g.][]{PL1997,ER2001,ER2003}. Nevertheless, it amounts to the torque exerted on the Earth's solid regions in order of magnitude \citep[e.g.][]{Lambeck1977,ML2009} although the atmosphere only represents ${\sim}\num{8.6e-5}\%$ of the planet's mass\footnote{The total mass of the atmosphere is $\num{5.136e18}$~kg while the Earth's mass is $\num{5.9723e24}$~kg \citep[e.g.][14-10]{Lide2008book}.}. Likewise, atmospheric and solid-body tides drive Venus towards an asynchronous rotation state of equilibrium by approximately counterbalancing each other \citep[e.g.][]{GS1969,ID1978,CL2001,Correia2003,CL2003,Leconte2015,Revol2023}. This mechanism is expected to determine the rotational state of near synchronous rocky exoplanets as well \citep[e.g.][]{LC2004proc,CL2010,Cunha2015,Leconte2015,ADLM2017a,ADLM2017b,Revol2023,SW2024}.

Furthermore, the atmospheric response strongly depends on the forcing frequency, analogous to ocean tides. This frequency-dependence is specific to planetary fluid layers and stars where, unlike solid regions, the phase velocities of propagating waves are close to those of tidal perturbations, thus leading to significant inertial effects. These forced waves form the so-called dynamical tide, while the quasi-static adjustment observed in solid regions and in the zero-frequency limit in general is referred to as the equilibrium tide \citep[e.g.][]{Zahn1975,Zahn1989,Ogilvie2014}. The tidal response of planetary atmospheres is a combination of several predominant types of waves arising from tidal gravitational forces and heating: inertial waves, internal gravity waves (in the presence of stable stratification), and Lamb waves, which are restored by Coriolis acceleration, the Archimedean force, and compressibility, respectively \citep[e.g.][]{Vallis2017}. Particularly, the latter play a key role in the rotational evolution of fast-rotating rocky planets, such as the Earth. 

Named after Lamb's pioneering work \citep[][]{Lamb1911}, Lamb waves are horizontally propagating planetary-scale waves similar to the long-wavelenth surface gravity modes of barotropic ocean tides \citep[][]{Bretherton1969,Lindzen1972}. They primarily account for the global atmospheric mass redistribution associated with thermal tides. Notably, the dominating Lamb wave raised by the semidiurnal tidal heating can be resonantly excited in the frequency range of typical tidal periods, thus significantly enhancing the associated thermotidal torque. \cite{ZW1987} noted that such a resonance occurred on Earth during the Precambrian Era ($540{-}4570$~Myr) when the length of the day (LOD)\mynom[A]{LOD}{length of the day} was approximately 21~hours, coinciding with a period of weaker ocean tides. This statement led them to hypothesise that the Earth's LOD could have been stalled for hundreds of million years by ocean and thermotidal torques compensating each other, like Venus' LOD, which would have significant implications for the Earth climatic history and, possibly, past oxygenation events \citep[e.g.][]{Jenkins1993,Klatt2021}. 

Whereas it received little attention for three decades, the role played by thermal tides in the history of the Earth's rotation has resurfaced in the 2010s \citep[][]{BS2016,Wu2023,Farhat2024,Laskar2024,DG2024}. This was both due to a growing amount of deep-time geological constraints \citep[e.g.][]{Zhou2024,Huang2024,Wu2024}, and to the observed mismatch between theoretical predictions solely based on oceanic tidal dissipation and Lunar rock datations for the age of the Moon \citep[e.g.][]{BR1999}. However, attempts to characterise the atmospheric contribution to long-term LOD variations have reached conflicting conclusions so far. Several authors argue for a billion-year LOD stalling around ${\sim} 19$~hours during the Proterozoic \citep[$540{-}2500$~Myr;][]{MK2023,Wu2023}, while others find that the Lamb wave resonance has likely never been strong enough to counterbalance the deceleration caused by ocean tides \citep[][]{Farhat2024,Laskar2024}. Incidentally, \cite{Farhat2022a} show that thermal tides are actually not required to construct a self-consistent theory of the Earth-Moon system's evolution that fits geological data. These discrepancies stem from the complexity of the problem, which requires rigorous evaluation of both the atmospheric and oceanic tidal torques.

Two radically different approaches are commonly used to quantify the thermotidal torque and to characterise its dependence on the tidal frequency. The first one builds on Laplace's masterpiece, which constitutes the foundation of modern modelling for planetary fluid tides \citep[][]{Laplace1798}. In this approach, the atmospheric response to a supposedly small tidal perturbation is analytically derived from the primitive equations of fluid dynamics, which are solved over a thin spherical shell in a linear framework \citep[e.g.][]{Wilkes1949,Siebert1961,LC1969}. This allows the tidal fields and torque to be expressed as explicit functions of the tidal frequency and a limited number of key physical parameters \citep[e.g.][]{LM1967,Lindzen1968,LB1972,ADLM2017a,Farhat2024}. Although they rely on substantial mathematical simplifications, analytical methods turn out to be well suited to the study of poorly constrained systems because they provide deep insights about the fundamental role played by the involved physical mechanisms. Besides, they enable the coherent calculation of the coupled tidal-orbital evolutions of planet-star systems including thermal tides \citep[e.g.][]{Revol2023,VA2023,Valente2024}. 

However, analytical models necessarily come with free parameters accounting for complex processes that cannot be self-consistently incorporated in closed-form solutions owing to the associated mathematical complications. Typically, the amplification of the tidal torque while crossing the Lamb wave resonance is determined by the efficiency of dissipative mechanisms, which is treated as an input in the analytical theory. In order to remedy to these limitations, several authors have elaborated numerical methods based on GCM simulations over the past decade \citep[e.g.][]{Leconte2015,ADL2019,Wu2023,DG2024,SW2024}. This second approach involves the time-integration of the 3D atmospheric dynamics taking into account its coupling with radiative transfer and dissipative mechanisms in an exhaustive way. Consequently, it yields valuable constraints for the free parameters of the analytical theory while allowing the validity of closed-form solutions to be tested. However, GCM simulations do not enable a wide exploration of the parameter space because they are computationally expensive. Finally, it is noteworthy that many authors have refined the classical tidal theory since the 1970s by developing 2D and 3D models of increasing complexities \citep[e.g.][]{LH1974,FG1979,Forbes1982,Vial1986,WA1997,Hagan1995,Hagan1999,Huang2007,FZ2022}, but these intermediate models do not seem to have been used yet to investigate the action of thermal tides on the rotation of rocky planets.

In the present work, we aim at bridging the gap between analytical and numerical solutions by showing quantitatively how they may converge or diverge from each other. Because this goal requires numerous derivations that could detract from the clarity of the discussion, we proceed in three steps. In this first article (Paper~I), we establish a fully analytical model describing the atmospheric tidal response of rocky planets in the framework of the classical theory \citep[e.g.][]{CL70}. This model is generic enough to be applicable to both the Earth and rocky extrasolar planets. Dissipative effects are incorporated using Newtonian cooling. 

The tidal fields (pressure, density, temperature, wind velocities) are derived for two idealised vertical profiles of background temperature framing the typical atmospheric structures of terrestrial planets. In the first configuration, which we call `isentropic', the atmosphere is neutrally stratified, meaning that internal gravity waves are filtered out. The other configuration is the standard isothermal atmosphere, where the background temperature is globally uniform. In a second paper (\papertwo; Auclair-Desrotour et al. 2026b), we will use this analytical model in both isentropic and isothermal configurations (i) to express the thermotidal torque exerted about the planet's spin axis as a function of the tidal frequency, (ii) to characterise analytically the principal features of its frequency dependence -- including the Lamb wave resonance --, and (iii) to benchmark the theory against GCM simulations. In a third paper (\paperthree;  Auclair-Desrotour et al. 2026c), we will present an equivalent mass-spring-damper model that accurately reproduces our analytic solution for thermotidal surface pressure oscillations and we will investigate the influence of indirect tidal heating on the resulting torque. 

In \sect{sec:tidal_dynamics}, we introduce the dissipative equations of tidal dynamics for arbitrary vertical profiles of background quantities, and we discuss the horizontal and vertical structures of tidal modes. In \sect{sec:solution}, we solve tidal equations analytically for isentropic and isothermal atmospheres, showing that these configurations are actually two particular cases of the same solution. Also, we formulate all the tidal fields as functions of this solution, and we recover in a generalised form the analytical expressions of Lamb waves resonance frequencies established in earlier studies. Finally, we discuss the model limitations in \sect{sec:model_limitations}, and we present our conclusions in \sect{sec:conclusions}. Notations and acronyms used throughout this paper are listed in the nomenclature in \append{app:nomenclature}.





\begin{figure}[t]
   \centering
   \includegraphics[width=0.48\textwidth,trim = 0.cm 0.cm 17.9cm 11.3cm,clip]{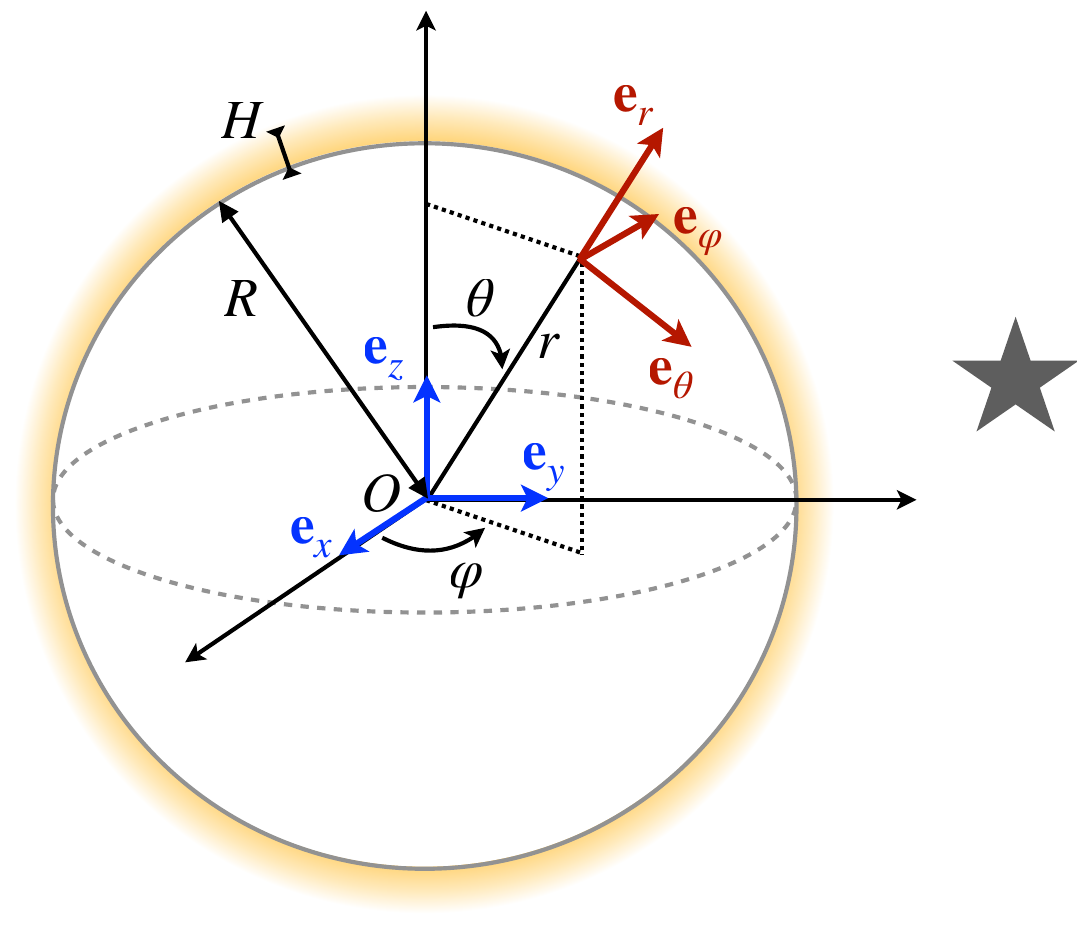}
      \caption{Frame of reference and system of coordinates. Blue arrows denote the Cartesian unit vectors $\left( \ex,\ey, \ez \right)$ associated with the frame of reference of the planet, $\framerot$, and red arrows the unit vector basis $\left( \er , \etheta, \ephi \right)$ associated with the standard spherical coordinates $\left( \rr  , \col , \lon \right)$, with $\rr$, $\col$, and $\lon$ being the radial coordinate, the colatitude, and the longitude, respectively. Also are shown the planet's centre of mass, $O$, which serves as the origin of $\framerot$, the planet radius, $\Rpla$, and the pressure height of the atmosphere, $\height \ll \Rpla$.  }
       \label{fig:diagram_frame}%
\end{figure}

\section{Tidal dynamics}
\label{sec:tidal_dynamics}
\subsection{Physical setup}
\label{ssec:physical_setup}

We study thermal tides in the framework of the classical tidal theory \citep{LC1969}, considering a rocky planet of radius $\Rpla$\mynom[S]{$\Rpla$}{planet radius} and surface gravity $\ggravi$\mynom[S]{$\ggravi$}{surface gravity}, as illustrated by \fig{fig:diagram_frame}. The planet's spin rotation is defined by the rotation vector $\spinvect$\mynom[S]{$\spinvect$}{rotation vector}. Assuming that the planet's axis of figure and axis of rotation are the same, we introduce the non-Galilean frame of reference of the planet, $\rframe{\irot}{O}{\ex}{\ey}{\ez}$\mynom[S]{$\framerot$}{non-Galilean frame of reference of the planet}, which has the planet's centre of mass, $O$\mynom[S]{$O$}{planet's centre of mass}, as origin. The Cartesian unit vectors $\ex$\mynom[S]{$\ex,\ey$}{orthogonal Cartesian unit vectors defining the planet's equatorial plane} and $\ey$ define two orthogonal direction of the planet's equatorial plane, while $\ez$ is aligned with the spin axis\mynom[S]{$\ez$}{Cartesian unit vector aligned with the spin axis}. The planet's spin vector is thus defined as $\spinvect = \spinrate \ez$, with $\spinrate$ being the planet's spin angular velocity\mynom[S]{$\spinrate$}{planet's spin angular velocity}. To define the position of any point in $\framerot$, we adopt spherical coordinates $\left( \rr , \col , \lon \right)$ centred on the planet's centre of mass, with $\rr$ being the radial coordinate\mynom[S]{$\rr$}{radial coordinate}, $\col$ the colatitude\mynom[S]{$\col$}{colatitude}, and $\lon$ the longitude\mynom[S]{$\lon$}{longitude}. These coordinates are associated with the usual basis unit vectors $\left( \er, \etheta, \ephi \right)$\mynom[S]{$\er$}{radial unit vector in $\framerot$}\mynom[S]{$\etheta$}{colatitudinal unit vector in $\framerot$}\mynom[S]{$\ephi$}{longitudinal unit vector in $\framerot$}, given by 
\begin{equation}
\begin{array}{l}
\er = \cos \col \, \ez + \sin \col \left( \cos \lon \, \ex + \sin \lon \, \ey \right) , \\
 \etheta = - \sin \col \, \ez + \cos \col \left( \cos \lon \, \ex + \sin \lon \, \ey \right) , \\
 \ephi= \cos \lon \, \ey - \sin \lon \, \ex. 
\end{array}
\end{equation}

The planet's atmosphere is assumed to be thin, meaning that its typical pressure height, $\height$\mynom[S]{$\height$}{pressure height}, is such that $\height \ll \Rpla$. This allows the gravity field's dependence on altitude to be neglected. Also, the centrifugal acceleration associated with the planet's spin rotation is ignored, so that $\ggravi$ is uniform in the atmospheric shell. Similarly, one replaces $\rr$ by $\Rpla$ in the metric factors of the fluid dynamical equations, thus disregarding curvature variations with altitude, $\zz$\mynom[S]{$\zz$}{altitude}. The planet's ellipticity and topography are ignored as well. Following usual simplifications, mean flows associated with general circulation are not taken into account and the atmospheric structure is assumed to be spherically symmetric about the planet's centre of mass. 

The atmosphere is finally treated as an ideal gas characterised solely by its specific gas constant, $\Rspec = \Rgp/ \Mmol$\mynom[S]{$\Rspec$}{specific gas constant} (with $\Rgp$\mynom[S]{$\Rgp$}{universal gas constant} and $\Mmol$\mynom[S]{$\height$}{molar mass of the atmospheric mixture} being the universal gas constant and the molar mass of the atmospheric mixture, respectively), and its adiabatic index, $\adiabexp  = \Cp / \Cv$\mynom[S]{$\adiabexp$}{adiabatic index} (with $\Cp$\mynom[S]{$\Cp$}{heat capacity of the gas at constant pressure} and $\Cv$\mynom[S]{$\Cv$}{heat capacity of the gas at constant volume} denoting the heat capacities at constant pressure and constant volume, respectively). The latter is greater than 1 and decreases as the size of gas molecules increases, since it is linked to the number of degrees of freedom of the atoms in molecules \citep[e.g.][Eqs.~(44.1) and~(44.2), p.~124]{LL1969book}. Values of $\adiabexp$ given by the kinetic theory of gases are typically 5/3 for monoatomic gases, and 7/5 for diatomic gases \citep[e.g.][Sect.~1.8]{Vallis2017}. More realistic values taking into account the temperature dependence of the adiabatic index for various gases can be found in \cite{Dean1973book}, Table~9.7, p.~9-116. Here, both $\Rspec$ and $\adiabexp$ are assumed to be spatially uniform and invariant in time.

Tides are treated as linearisable perturbations about a basic state. The local pressure, $\press$\mynom[S]{$\press$}{pressure}, density, $\density$\mynom[S]{$\density$}{density}, and temperature, $\temp$\mynom[S]{$\temp$}{temperature}, are decomposed into background fields and tidal fluctuations\mynom[S]{$\pressbg$}{background pressure}\mynom[S]{$\rhobg$}{background density}\mynom[S]{$\tempbg$}{background temperature}\mynom[S]{$\presstide$}{tidal pressure fluctuation}\mynom[S]{$\rhotide$}{tidal density fluctuation}\mynom[S]{$\temptide$}{tidal temperature fluctuation},
\begin{align}
\label{press_field}
\press  & = \pressbg \left(\zz\right)  + \presstide \left( \col , \lon, \zz , \time \right) , \\
\density  & = \rhobg \left(\zz\right)  + \rhotide \left( \col , \lon, \zz , \time \right) , \\
\temp  & = \tempbg \left(\zz\right)  + \temptide \left( \col , \lon, \zz , \time \right),
\end{align}
which are complemented by the velocity field of tidal flows\mynom[S]{$\Vvect$}{velocity vector}\mynom[S]{$\Vr$}{radial velocity}\mynom[S]{$\Vtheta$}{colatitudinal velocity}\mynom[S]{$\Vphi$}{longitudinal velocity},
\begin{equation}
\label{velocity_vector}
\Vvect = \Vr \left( \col , \lon, \zz, \time \right) \er +  \Vtheta \left( \col , \lon, \zz , \time \right) \etheta + \Vphi \left( \col , \lon, \zz , \time \right) \ephi.
\end{equation} 
The background fields -- subscripted by~$0$ in the above equations -- only vary with the vertical coordinate. The tidal fields are referred to by the symbol $\tidal{}$, except velocities. Unlike background fields, they are functions of spatial coordinates and time,~$\time$\mynom[S]{$\time$}{time}. 

Considering that vertical accelerations are small compared to gravity and pressure forces, we assume that the air column is in hydrostatic equilibrium. It follows\mynom[S]{$\DD{}{X}$}{material derivative with respect to $X$}
\begin{equation}
\DD{\pressbg}{\zz} = - \ggravi \rhobg. 
\end{equation}
This allows for defining the altitude-dependent pressure height as 
\begin{equation}
\label{pressure_height}
\height \left( \zz \right) = \frac{\pressbg}{\ggravi \rhobg},
\end{equation}
and the nondimensional pressure coordinate $\xx$\mynom[S]{$\xx$}{nondimensional pressure coordinate} as 
\begin{equation}
\label{vertical_coord}
\xx = \int_0^\zz \frac{\infvar{\zz^\prime}}{\height \left( \zz^\prime \right)},
\end{equation}
with $\infvar{\zz^\prime}$ being an infinitesimal altitude difference. We note that $\xx$ is also given by $\xx = - \ln \left( \pressbg / \psurf \right)$, where $\psurf$ designates surface pressure\mynom[S]{$\psurf$}{globally averaged surface pressure}. Therefore, all equations of the present study can be rewritten in pressure coordinates $\xp = \pressbg/\psurf$\mynom[S]{$\xp$}{standard pressure coordinate} by substituting the vertical gradient of any field $f$ with $\dd{f}{\xx} = - \xp \dd{f}{\xp}$\mynom[S]{$\dd{}{X}$}{partial derivative with respect to $X$}. Besides, using the ideal gas law,
\begin{equation}
\pressbg = \Rspec  \rhobg \tempbg,
\end{equation}
the pressure height defined in \eq{pressure_height} can be reformulated as a function of temperature,
\begin{equation}
\label{pressure_height_temp}
\height = \frac{\Rspec \tempbg}{\ggravi}.
\end{equation}

The atmospheric stability with respect to convection is quantified by the Brunt-Väisälä frequency, $\BVfreq$\mynom[S]{$\BVfreq$}{Brunt-Väisälä frequency} \citep[][Chap.~1, Sect.~4]{Unno1989}, which is given, in the hydrostatic approximation, by
\begin{equation}
\label{BVfreq}
\BVfreq^2 = \frac{\ggravi}{\height} \left( \rcp + \DD{\ln \height}{\xx} \right),
\end{equation}
with $\rcp = \left( \adiabexp - 1 \right) / \adiabexp$\mynom[S]{$\rcp$}{$\rcp = \left( \adiabexp - 1 \right) / \adiabexp$} \citep[the kinetic theory of gases yields $\rcp = 2/5 =0.4$ for monoatomic gases, $\rcp = 2/7 \approx 0.286$ for diatomic gases; see e.g.][Sect.~1.8]{Vallis2017}. The Brunt-Väisälä frequency is the typical frequency with which a bubble of gas may oscillate vertically around its equilibrium position under gravity. If $\BVfreq>0$, the atmosphere is stably stratified, meaning that vertical oscillations of fluid particles are restored by the Archimedean force, as observed in stellar radiative regions and in the Earth's stratosphere. Conversely, if $\BVfreq=0$, the temperature profile is neutrally stratified and the Archimedean force vanishes. This temperature profile typically corresponds to convective regions in fluid layers. It is commonly named `isentropic' or `adiabatic' given that entropy is conserved across the vertical direction \citep[i.e. the potential temperature is uniform, see e.g.][]{Pierrehumbert2010}. Particularly, the isentropic profile approximately describes the atmospheric structure of Earth's troposphere, which is the layer extending from the ground to a pressure level of ${\sim}100$~mb \citep[e.g.][Chap.~2, Sect.~2.2]{Pierrehumbert2010}.  

In the following, we establish analytical solutions for the atmospheric tidal response of two types of atmospheres that have been used since the foundations of tidal theory \citep[][]{Lamb1932book,Siebert1961} : (i) strictly isentropic atmospheres, such that $\BVfreq=0$ over the whole air column ; and (ii) isothermal atmospheres, such that $\tempbg $ is globally uniform. To some extent, these two configurations are end-member scenarios for the atmospheric structure. On the one hand, the isentropic -- or autobarotropic \citep[e.g.][]{Siebert1961} -- profile is representative of atmospheres in convective equilibrium. Remarkably, tidal oscillations of oceans and isentropic atmospheres are completely analogous \citep[][]{Bartels1927}. The isentropic setup thus seems appropriate to investigate analytically the implications of thermal tides on Earth's rotation, as most of the mass redistribution inducing the tidal torque occurs within the troposphere, which contains ${\sim}80\%$ of the atmospheric mass \citep[e.g.][Sect.~15.5]{Vallis2017}. On the other hand, the isothermal scenario can be considered as a simplified model for strongly stratified atmospheres. In such atmospheres, tidally forced internal gravity waves -- namely waves restored by the Archimedean force -- can propagate upwards starting from the planet's surface, as observed on Pluto \citep[e.g.][]{Toigo2010,Gladstone2016,Jacobs2021}. Comparing the tidal solutions obtained in the two scenarios provides insights about the sensitivity of thermal tides to atmospheric structure.

\subsection{Thermo-gravitationally forced tidal waves}

Thermal tides are generated by external heating sources through the thermotidal heating per unit mass per unit time, $\inputpow$\mynom[S]{$\inputpow$}{thermotidal heating per unit mass per unit time}, which is the notation classically adopted in literature \citep[e.g.][]{Siebert1961,CL70,ADLM2017a}. This term enters in the thermodynamic equation, given by \eq{thermo_equation}, and represents the time-varying component of the power absorbed locally by the gas per unit mass. Similarly, the forces~$\forcev$ responsible for gravitational tides are given by $\forcev = \grad \gravpot$\mynom[S]{$\forcev$}{gravitational tidal forces}\mynom[S]{$\grad$}{gradient operator}, with $\gravpot$\mynom[S]{$\gravpot$}{tidal gravitational potential} denoting the tidal gravitational potential\footnote{Here we adopt Zahn's convention for the definition of the tidal potential \citep[see e.g.][]{Zahn1966a}. This convention differs to a minus sign from that used by \cite{LC1969}.}. 
In linear theory, further simplifications are needed in order to allow the tidal problem to be handled analytically. First, we adopt the so-called Cowling approximation, meaning that we ignore the feedback of tidally induced self-attraction variations on the atmospheric tidal dynamics \citep[][]{cowling1941,Unno1989}. Then, we assume the `traditional approximation' (TA). This implies neglecting the horizontal components of Coriolis acceleration induced by vertical motions, as well as the vertical components induced by horizontal motions \citep[e.g.][Chap.~6, Sect.~34.3]{Unno1989}. 

As it is highlighted by \cite{Gerkema2008}, the justification for the TA in the thin-shell approximation is mainly based on the net separation of length scales between vertical and horizontal dimensions. Since $\height \ll \Rpla$, the velocity components are such that $\abs{\Vr} \ll \abs{\Vtheta} , \abs{\Vphi}$, rendering the horizontal components of Coriolis accelerations associated with vertical fluid motions insignificant compared to those associated with horizontal motions. This argument, however, cannot be invoked to discard the vertical Coriolis acceleration induced by horizontal motions since the latter are significant. To get rid of this acceleration in the thin-shell approximation, the tidally perturbed pressure field can simply be taken as nearly hydrostatic, similar to the equilibrium pressure field. Under this approximation, all non-hydrostatic terms -- including Coriolis acceleration -- are negligible compared to pressure forces and buoyancy, defined as $\buoyancy = - \ggravi \rhotide / \rhobg$\mynom[S]{$\buoyancy$}{buoyancy} \citep[][]{GZ2008}. 

In non-hydrostatic or thick-shell frameworks, the TA requires assuming strong vertical stratification in density, which suppresses large-scale vertical motions \citep[][]{Unno1989,GS2005,Gerkema2008}. The TA is thus usually considered as an appropriate simplification in stably stratified fluid layers such as the isothermal atmosphere of the second scenario, where Archimedean forces drive the vertical component of tidal wave dynamics. 
However, one may legitimately question the significance of buoyancy in the absence of stable stratification. The fact that there is no restoring force across the vertical direction in this configuration suggests that $\buoyancy$ should be negligible. As a consequence, ignoring vertical Coriolis accelerations might seem contradictory with the isentropic atmosphere hypothesis ($\BVfreq=0$). 

This paradox can actually be resolved by noticing that the `buoyancy' terminology is misleading because it eludes density variations caused by horizontal compressibility. Compressibility forces, far from being negligible, primarily control the global atmospheric tidal response, as it will be shown further. The associated horizontally propagating planetary-scale acoustic waves are called Lamb waves \citep[e.g.][]{Bretherton1969,Lindzen1972}. These waves are analogous to the long-wavelength surface gravity modes that predominate in Earth's ocean tides \citep[][]{Bartels1927}. Consequently, buoyancy can be significant even in the absence of Archimedean forces and the TA still holds in this case. Moreover, it is noteworthy that hydrostatic balance is often assumed for the vertical momentum equation in rocky planet-oriented general circulation models (GCMs) regardless of atmospheric stability \citep[see, for example, the LMDZ GCM;][]{Hourdin2006}. 
The above simplifications are thus also consistent with the physical setup of numerical solutions computed from GCM simulations \citep[e.g.][]{Leconte2015,ADL2019,Wu2023,DG2024}. We refer the reader to \cite{Gerkema2008} for a thorough discussion about the limitations of the traditional approximation in fluid dynamics and fluid tides.

The primitive equations governing the tidal response of a thin planetary atmosphere are given by \citep[e.g.][]{LC1969}
\begin{align}
\label{momentum_theta}
\dd{\Vtheta}{\time} - 2 \spinrate \cos \col \Vphi  + \frac{1}{\Rpla} \dd{}{\col} \left( \frac{\presstide}{\rhobg}  - \gravpot \right) = 0, \\
\label{momentum_phi}
\dd{\Vphi}{\time} + 2 \spinrate \cos \col \Vtheta  + \frac{1}{\Rpla \sin \col} \dd{}{\lon} \left( \frac{\presstide}{\rhobg}  - \gravpot \right) =0, \\
\label{momentum_vert}
\dd{\presstide}{\xx} + \ggravi \height \rhotide  - \rhobg \dd{\gravpot}{\xx} = 0, \\
\label{mass_conservation}
\DD{\density}{\time} + \rhobg \left(\frac{1}{\height} \dd{\Vr}{\xx} + \divh \Vvect \right) = 0, \\
\label{thermo_equation}
\DD{\press}{\time} - \csound^2 \DD{\density}{\time}  -  \left( \adiabexp - 1 \right) \rhobg \inputpow + \rhobg \Rspec  \adiabexp \fnewton \temptide =0,   \\
\label{ideal_gas_equation}
\frac{\presstide}{\pressbg} - \frac{\temptide}{\tempbg} - \frac{\rhotide}{\rhobg} = 0,
\end{align}
where we have introduced the sound speed\mynom[S]{$\csound$}{sound speed},
\begin{equation}
\csound = \sqrt{\adiabexp \ggravi \height}.
\end{equation}
In the above system of equations, \eqsthree{momentum_theta}{momentum_phi}{momentum_vert} are the $\col$, $\lon$ and $\zz$-components of the momentum equation, respectively; \eq{mass_conservation} is the mass conservation equation; \eq{thermo_equation} is the thermodynamic equation; and \eq{ideal_gas_equation} is the linearised ideal gas law. The notation $\divh \Vvect$\mynom[S]{$\divh$}{horizontal divergence operator} in \eq{mass_conservation} designates the horizontal divergence of the velocity vector,
\begin{equation}
\label{divhV}
\divh \Vvect = \frac{1}{\Rpla \sin \col} \left[ \dd{}{\col} \left( \sin \col \Vtheta \right) + \dd{\Vphi}{\lon} \right],
\end{equation}
with $\dd{}{X}$ being the partial derivative operator with respect to any coordinate $X$. The material derivative $\DD{\yy}{\time}$ of any field $\yy$ given by \eqsto{press_field}{velocity_vector}, is expressed as 
\begin{equation}
\DD{\yy}{\time} = \dd{\tidal{\yy}}{\time} + \frac{\Vr}{\height} \DD{\yy_0}{\xx},
\end{equation}
where $\yy_0$\mynom[S]{$\yy_0$}{background field of $\yy$} and $\tidal{\yy}$\mynom[S]{$\tidal{\yy}$}{tidal fluctuation of $\yy$} designate the background and tidal fields of $\yy$, respectively, and $\xx$ is the nondimensional pressure coordinate introduced in \eq{vertical_coord}. 

We emphasise that the thermodynamic equation given by \eq{thermo_equation} slightly differs from that used in the classical theory of atmospheric tides \citep[see e.g.][Chap.~3, Eq.~(13)]{LC1969}. Following \cite{LM1967} and \cite{Lindzen1968application}, we have included Newtonian cooling in order to account for the effects of dissipative processes on tidal dynamics. This term represents any energy loss scaling linearly with temperature fluctuations, such as radiative cooling of fluid particles towards space. It is parametrised by the cooling frequency, $\fnewton$\mynom[S]{$\fnewton$}{Newtonian cooling frequency}, which controls the efficiency of dissipative mechanisms. All along \sect{sec:tidal_dynamics}, $\fnewton$ is assumed to be a function of altitude like other background quantities for the sake of generality. 


\subsection{Tidal equations in frequency domain}

As the tidal oscillation is periodic both in time and longitude, it is convenient to rewrite \eqsto{momentum_theta}{ideal_gas_equation} in the frequency domain. Each Fourier component can be considered independently of the others because of linearity. We thus focus on one component of tidal frequency $\ftide$\mynom[S]{$\ftide$}{tidal frequency} and order $\mm$\mynom[S]{$\mm$}{integral order of the spherical harmonics} in the following, and we ignore all the other components. In order to emphasise the dimensionless parameters that control tidal dynamics, we formulate tidal equations in a nondimensional form. This is done by introducing the normalised time $\timen = \ftide \time$\mynom[S]{$\timen$}{normalised time}, the reference velocity $\speedref = \Rpla \ftide$\mynom[S]{$\speedref$}{reference velocity}, and the dimensionless spatial distributions $\Vthetan$\mynom[S]{$\Vthetan$}{normalised colatitudinal-velocity field for the ($\ftide, \mm$) component}, $\Vphin$\mynom[S]{$\Vphin$}{normalised longitudinal-velocity field for the ($\ftide, \mm$) component}, $\Vrn$\mynom[S]{$\Vrn$}{normalised radial-velocity field for the ($\ftide, \mm$) component}, $\presstiden$\mynom[S]{$\presstiden$}{normalised pressure field for the ($\ftide, \mm$) component}, $\rhotiden$\mynom[S]{$\rhotiden$}{normalised density field for the ($\ftide, \mm$) component}, $\temptiden$\mynom[S]{$\temptiden$}{normalised temperature field for the ($\ftide, \mm$) component}, $\Jtiden$\mynom[S]{$\Jtiden$}{normalised field of tidal heating for the ($\ftide, \mm$) component} and $\gravpotn$\mynom[S]{$\gravpotn$}{normalised field of tidal gravitational potential for the ($\ftide, \mm$) component} such that 
\begin{align}
\label{Vtheta_Hough}
\Vtheta \left( \xx, \col, \lon, \timen \right) & = \Re \left\{ \speedref  \Vthetan \left( \xx, \col \right) \expo{\inumber \left( \timen + \mm \lon \right)} \right\} ,\\ 
\Vphi \left( \xx, \col, \lon, \timen \right)  & = \Re \left\{ \speedref \Vthetan \left( \xx, \col \right) \expo{\inumber \left( \timen + \mm \lon \right)} \right\} , \\
\Vr \left( \xx, \col, \lon, \timen \right)  & = \Re \left\{ \ftide \height \left( \xx \right) \Vthetan \left( \xx, \col \right) \expo{\inumber \left( \timen + \mm \lon \right)} \right\} , \\
\label{presstide_Hough}
\presstide \left( \xx, \col, \lon, \timen \right) & = \Re \left\{ \pressbg \left( \xx \right)  \presstiden \left( \xx, \col \right) \expo{\inumber \left( \timen + \mm \lon \right)} \right\} ,\\ 
\rhotide \left( \xx, \col, \lon, \timen \right) & = \Re \left\{ \rhobg \left( \xx \right)  \rhotiden \left( \xx, \col \right) \expo{\inumber \left( \timen + \mm \lon \right)} \right\} ,\\ 
\label{temptide_Hough}
\temptide \left( \xx, \col, \lon, \timen \right) & = \Re \left\{ \tempbg \left( \xx \right)  \temptiden \left( \xx, \col \right) \expo{\inumber \left( \timen + \mm \lon \right)} \right\} ,\\ 
\Jtide \left( \xx, \col, \lon, \timen \right) & = \Re \left\{ \ftide \rcp^{-1} \ggravi \height \left( \xx \right) \Jtiden \left( \xx, \col \right) \expo{\inumber \left( \timen + \mm \lon \right)} \right\} ,\\ 
\Utide \left( \xx, \col, \lon, \timen \right) & = \Re \left\{ \ggravi \height \left( \xx \right) \Utiden \left( \xx, \col \right) \expo{\inumber \left( \timen + \mm \lon \right)} \right\} ,
\end{align}
where $\Re$\mynom[S]{$\Re$}{real part of a complex number} refers to the real part of a complex number ($\Im$\mynom[S]{$\Im$}{imaginary part of a complex number} referring to the imaginary part), and $\inumber$\mynom[S]{$\inumber$}{imaginary unit} is the imaginary unit. Additionally, we denote by $\divhn$\mynom[S]{$\divhn$}{normalised horizontal divergence operator} the normalised horizontal divergence operator, defined from \eq{divhV} as 
\begin{equation}
\divhn = \Rpla  \divh,
\end{equation}
and we introduce the dimensionless quantities
\begin{equation}
\begin{array}{lllll}
\displaystyle \spinpar = \frac{2 \spinrate}{\ftide}, & 
\displaystyle \alphapar  = \frac{\fnewton}{\ftide}, & 
\displaystyle \betapar= \frac{\ggravi \height}{\speedref^2}, & 
\displaystyle \gammapar  = \frac{\BVfreq^2 \height}{\ggravi}, &
\displaystyle \dlnbet = \DD{\ln \betapar}{\xx}.
\end{array}
\label{dimensionless_numbers}
\end{equation}

The notation $\spinpar$\mynom[S]{$\spinpar$}{spin parameter} designates the spin parameter, which compares Coriolis acceleration to the local acceleration of a fluid particle \citep[e.g.][]{LS1997}. If $\abs{\spinpar} \ll 1$, Coriolis acceleration has a negligible impact on tidal dynamics, while it strongly distorts tidal flows if $\abs{\spinpar} \gg 1$. Similarly, $\alphapar$\mynom[S]{$\alphapar$}{dimensionless radiative relaxation parameter} measures the influence of radiative relaxation on tidal dynamics. If $\alphapar \ll 1$, the relaxation is slow and the tides are just weakly attenuated by radiative cooling. If $\alphapar \gg 1$, the relaxation is fast, which strongly damps tidal oscillations. The dimensionless number $\betapar$\mynom[S]{$\betapar$}{dimensionless number weighting gravitational effects relative to inertial ones} quantifies gravitational effects relative to inertial ones. It corresponds to the inverse of a squared Froude number \citep[e.g.][]{Vallis2017}. If $\betapar \ll 1$, the atmospheric tidal response is driven by inertial terms, while the atmosphere hydrostatically adjusts to the tidal forcing if $\betapar \gg 1$. 

The nondimensional quantity $\gammapar$\mynom[S]{$\gammapar$}{reduced stratification number} is a reduced stratification number accounting for the influence of Archimedean forces on tidal dynamics. If $\gammapar =0$, the atmosphere is unstratified and a fluid particle can freely move across the vertical direction without being affected by Archimedean forces. If $\gammapar >0 $, the Archimedean forces make particles oscillate around an equilibrium position. Finally, the dimensionless number $\dlnbet$\mynom[S]{$\dlnbet$}{reduced vertical variation rate characterising the atmospheric structure} is a reduced vertical variation rate characterising the atmospheric structure. This variation rate is directly related to the local vertical temperature gradient. In an isothermal atmosphere, the pressure height is uniform over the air column, which implies $\dlnbet = 0$. In an isentropic atmosphere, the temperature gradient is negative and \eq{BVfreq} leads to $ \dlnbet=- \rcp$. It is noteworthy that $\alphapar$, $\betapar$, $\gammapar$, and $\dlnbet$ are functions of the vertical coordinate in the general case. 

Using \eqsto{Vtheta_Hough}{dimensionless_numbers} in the horizontal momentum equations (\eqs{momentum_theta}{momentum_phi}), we obtain
\begin{align}
\label{Vtheta_deltap}
\Vthetan &=  \inumber \betapar  \Lapcol \left( \presstiden - \Utiden \right), \\
\label{Vphi_deltap}
\Vphin & =  - \betapar \Laplon \left( \presstiden - \Utiden  \right).
\end{align}
where $\Lapcol$\mynom[S]{$\Lapcol$}{operator defining the horizontal structure of $\Vthetan$} and $\Laplon$\mynom[S]{$\Laplon$}{operator defining the horizontal structure of $\Vphin$} are two operators acting on the $\col$-dependent part of the tidal perturbation, defined as \citep[e.g.][]{ADL2018}
\begin{align}
\label{lapcol}
\Lapcol &= \frac{1}{1- \spinpar^2 \cos^2 \col} \left( \dd{}{\col} + \spinpar \, \mm \, \cotan \col \right), \\
\label{laplon}
\Laplon & =   \frac{1}{1- \spinpar^2 \cos^2 \col} \left( \spinpar \cos \col \dd{}{\col} + \frac{\mm}{\sin \col} \right) . 
\end{align}
We then use \eqs{Vtheta_deltap}{Vphi_deltap} in \eq{divhV}. It follows
\begin{equation}
\label{divV_laplace}
\divhn \Vvectn = \inumber \betapar \Laplace \left( \presstiden - \Utiden \right),
\end{equation}
where $\Laplace$\mynom[S]{$\Laplace$}{Laplace tidal operator} is the Laplace tidal operator \citep[e.g.][]{Wang2016}, defined as
\begin{align}
\label{Laplace_operator}
\Laplace = & \frac{1}{\sin \col} \dd{}{\col} \left( \frac{\sin \col}{1 - \spinpar^2 \cos \col^2} \dd{}{\col} \right) \\
   & - \frac{1}{1 - \spinpar^2 \cos^2 \col} \left(  \mm \spinpar \frac{1 + \spinpar^2 \cos^2 \col}{1 - \spinpar^2 \cos^2 \col} + \frac{\mm^2}{\sin^2 \col} \right) .
\end{align}
Similarly as $\Lapcol$ and $\Laplon$, this operator only acts on the $\col$-dependent component of the tidal perturbation. 

Now, we formulate the system of equations given by \eqsto{momentum_theta}{ideal_gas_equation} as a single partial differential equation, which requires some additional manipulations. First, following \cite{LC1969}, we introduce the calculation variable $\Gtide$\mynom[S]{$\Gtide$}{calculation variable}, defined as
\begin{equation}
\label{eq_Gtide}
\Gtide = - \frac{1}{\adiabexp \pressbg} \DD{\press}{\time},
\end{equation}
and the corresponding nondimensional spatial distribution, $\Gtiden$\mynom[S]{$\Gtiden$}{normalised spatial distribution of $\Gtide$}, such that 
\begin{equation}
\Gtide \left( \xx, \col, \lon, \timen \right)  = \Re \left\{ \frac{\ftide}{\adiabexp} \Gtiden \left( \xx, \col \right) \expo{\inumber \left( \timen + \mm \lon \right)} \right\} .
\end{equation}
Then, we work out a compact form of \eqsto{momentum_vert}{thermo_equation} and \eq{eq_Gtide} in the frequency domain by using \eq{divV_laplace} in the equation of mass conservation and \eq{ideal_gas_equation} in the thermodynamic equation. This yields 
\begin{align}
\label{reduced_eq1}
\inumber \presstiden - \Vrn + \Gtiden = 0 , \\
\label{reduced_eq2}
\left( \dd{}{\xx} - 1 \right) \presstiden + \rhotiden - \left( \dd{}{\xx} + \dlnbet \right) \Utiden = 0 , \\
\label{reduced_eq3}
\rhotiden - \inumber \left( \dd{}{\xx} - 1 \right) \Vrn + \betapar \Laplace \left( \presstiden - \Utiden \right) = 0 , \\ 
\label{reduced_eq4}
\frac{\Gtiden}{\adiabexp} + \inumber \rhotiden - \left( 1 + \dlnbet \right) \Vrn + \Jtiden - \alphapar \temptiden = 0. 
\end{align}

A few steps only are needed to reduce these four equations to two partial differential equations for $\Gtiden$ and $\presstiden$. For the first equation, we express $\presstiden$ as a function of $\Gtiden$ and $\Vrn$ using \eq{reduced_eq1}, and we express $\rhotiden$ as a function of $\presstiden$, $\Vrn$ and $\Utiden$ using \eq{reduced_eq3}. Finally we substitute $\presstiden$ and $\rhotiden$ with the obtained expressions in \eq{reduced_eq2}, and we remark that all the terms involving vertical velocity vanish, which yields 
\begin{align}
\label{pde_1}
\dd{\Gtiden}{\xx} = \Gtiden  - \inumber  \betapar \Laplace \presstiden + \inumber \left( \betapar \Laplace - \dd{}{\xx} - \dlnbet \right) \Utiden.
\end{align}
For the second equation, we use the vertical momentum equation (\eq{reduced_eq2}) to express $\rhotiden$ as a function of $\presstiden $ and $\Utiden$, we use \eq{reduced_eq1} to express $\Vrn$ as a function of $\presstiden $ and $\Gtiden$, and we substitute $\rhotiden$ and $\Vrn$ with their expressions in \eq{reduced_eq4}. It follows
\begin{align}
\label{pde_2}
\dd{\presstiden}{\xx} = \left( 1 - \inumber \alphapar \right)^{-1} \left[ \inumber \gammapar \Gtiden -  \dlnbet \presstiden - \inumber \Jtiden \right] + \left( \dd{}{\xx} + \dlnbet \right) \Utiden.
\end{align}

As a last step, we apply the partial derivative operator $\dd{}{\xx}$ to \eq{pde_1} and we substitute $\dd{}{\xx}  \presstiden $ with \eq{pde_2} in the resulting equation. Noting that the Laplace tidal operator can be permuted with vertical structure quantities and operators, we reduce \eqsto{reduced_eq1}{reduced_eq4} to a single equation for $\Gtiden$ alone,
\begin{align}
\label{pdeG_single}
\ddd{\Gtiden}{\xx}{\xx} & - \dd{\Gtiden}{\xx} + \inumber \dd{}{\xx} \left[ \left( \dd{}{\xx} + \dlnbet\right) \Utiden \right] \\
& + \frac{\inumber \alphapar}{1 - \inumber \alphapar} \dlnbet \left[ \left( \dd{}{\xx} - 1 \right) \Gtiden + \inumber \left( \dd{}{\xx} + \dlnbet \right) \Utiden  \right] \nonumber \\
& = \frac{\betapar}{1 - \inumber \alphapar} \Laplace \left( \gammapar \Gtiden - \Jtiden - \alphapar \dlnbet \Utiden \right). \nonumber
\end{align}
This equation is analogous to Eq~(22) in \cite{LC1969}, Chap.~3, which can be recovered by setting $\alphapar$ to zero\footnote{There is a typo in Eq.~(22) of \cite{LC1969}: the factor of the third term in the left-hand member of the equation should read $\inumber \ftide / \left( \adiabexp \ggravi  \right) $ instead of $\inumber \ftide / \ggravi$.}. It shows that the vertical and horizontal structure components can be separated from each other, the former corresponding to the left-hand member of \eq{pdeG_single} and the latter to the right-hand member. Since $\Laplace$ admits a complete set of orthogonal eigenvectors for $0 \leq \col \leq \pi$, as discussed in \sect{ssec:horizontal_structure}, \eq{pdeG_single} may be solved by the method of separation of variables. Thus, the complex tidal fields are sought as series of the form
\begin{equation}
\label{hough_series}
\begin{array}{ll}
\displaystyle \Vthetan = \sum_\nn \Vthetav \left( \xx \right) \Houghcol{\nn}{\mm}{\spinpar} \left( \cos \col \right), &
\displaystyle \Vphin = \sum_\nn \Vphiv \left( \xx \right) \Houghlon{\nn}{\mm}{\spinpar} \left( \cos \col \right), \\
\displaystyle \Vrn = \sum_\nn \Vrv \left( \xx \right)  \HoughF{\nn}{\mm}{\spinpar} \left(\cos \col \right), & 
\displaystyle \presstiden = \sum_\nn \presstidev \left( \xx \right)  \HoughF{\nn}{\mm}{\spinpar} \left( \cos \col \right), \\ 
\displaystyle \rhotiden = \sum_\nn \rhotidev \left( \xx \right)  \HoughF{\nn}{\mm}{\spinpar} \left( \cos \col \right),&
\displaystyle \temptiden = \sum_\nn \temptidev \left( \xx \right)  \HoughF{\nn}{\mm}{\spinpar} \left(\cos  \col \right), \\
\displaystyle \Jtiden = \sum_\nn \Jtidev \left( \xx \right)  \HoughF{\nn}{\mm}{\spinpar} \left( \cos \col \right), & 
\displaystyle \Utiden = \sum_\nn \Utidev \left( \xx \right)  \HoughF{\nn}{\mm}{\spinpar} \left( \cos \col \right), \\
\displaystyle \Gtiden = \sum_\nn \Gtidev \left( \xx \right) \HoughF{\nn}{\mm}{\spinpar} \left( \cos \col \right), & 
\end{array}
\end{equation}
with $\nn$ being an integer\mynom[S]{$\nn$}{integral degree of Hough functions}, and 
\begin{align}
\label{hough_velocity}
& \Houghcol{\nn}{\mm}{\spinpar} = \Lapcol \HoughF{\nn}{\mm}{\spinpar}, & \Houghlon{\nn}{\mm}{\spinpar} = \Laplon \HoughF{\nn}{\mm}{\spinpar}.
\end{align}
The functions $\HoughF{\nn}{\mm}{\spinpar}$\mynom[S]{$\HoughF{\nn}{\mm}{\spinpar}$}{Hough function associated with the triplet $\left( \nn , \mm , \spinpar \right)$} introduced in \eq{hough_series} are named Hough functions after Hough's pioneering work \citep[][]{Hough1898}. They describe the horizontal structure of forced tidal modes. However, it is noteworthy that the horizontal distributions of tidal velocities are shaped by specific functions that differ from Hough functions, $\Houghcol{\nn}{\mm}{\spinpar}$\mynom[S]{$\Houghcol{\nn}{\mm}{\spinpar}$}{colatitudinal-velocity function associated with the triplet $\left( \nn, \mm , \spinpar \right)$} and $\Houghlon{\nn}{\mm}{\spinpar}$\mynom[S]{$\Houghlon{\nn}{\mm}{\spinpar}$}{longitudinal-velocity function associated with the triplet~$\left( \nn , \mm , \spinpar \right)$}. Analogously, the functions of $\xx$ in \eq{hough_series} describe the vertical structure of the modes. We emphasise that these functions are parametrised by $\mm$ and $\ftide$ like Hough functions. However we have dropped the superscripts in order to lighten expressions, keeping only the subscript,~$\nn$.

\subsection{Horizontal structure of tidal modes}
\label{ssec:horizontal_structure}

The horizontal structure of tidal waves has been thoroughly explored by many authors \citep[e.g.][]{LH1968,Holl1970,LS1997,Townsend2003,Wang2016}. In this section, we essentially refer to \cite{LS1997} as regards the main properties of the tidally forced modes. For a given doublet $\left( \mm , \spinpar \right)$, the Hough functions are the solutions of the eigenvalues-eigenvectors problem defined by 
\begin{equation}
\label{equation_laplace}
\left( \Laplace + \eigenval \right) \Hough = 0, 
\end{equation}
with regularity conditions applied at the two poles \citep[namely $\Hough \left( \col \right) $ is bounded at $\col = 0,\pi$; e.g.][]{LC1969}. Similarly as the associated Legendre functions, they form a set of orthogonal basis functions for the space of functions of $\col$ defined for $0 \leq \col \leq \pi$ \citep[][]{Holl1970,Wang2016}. Here, orthogonality is defined with respect to the scalar product given by 
\begin{equation}
\scal{F}{G} = \integ{F \left( \cos \col \right) G \left( \cos \col \right) \sin \col \,}{\col}{0}{\pi},
\end{equation}
with $F$ and $G$ being two functions of $\cos \col$ defined for $0 \leq \col \leq \pi$. Adopting the same normalisation as \cite{LC1969}, we use functions satisfying the condition
\begin{equation}
\scal{\HoughF{\nn}{\mm}{\spinpar}}{ \HoughF{\kk}{\mm}{\spinpar}} = \kron{\nn}{\kk},
\label{norm_cond_hough}
\end{equation}
with $\kron{\nn}{\kk}$\mynom[S]{$\kron{\nn}{\kk}$}{Kronecker delta function} denoting the Kronecker delta function (if $\nn=\kk$, then $\kron{\nn}{\kk}=1$, else $\kron{\nn}{\kk}=0$). This normalisation is similar to that of normalised associated Legendre functions \citep[ALFs, hereafter; see e.g.][Appendix~A]{Auclair2019}\mynom[A]{ALF}{normalised associated Legendre function}, as the degree-$\nn$ and $\kk$ ALFs sharing the same order $\mm$ -- denoted $\LegFnorm{\nn}{\mm}$\mynom[S]{$\LegFnorm{\nn}{\mm}$}{degree-$\nn$, order-$\mm$ ALF} and $\LegFnorm{\kk}{\mm}$, respectively -- are such that 
\begin{equation}
\scal{\LegFnorm{\nn}{\mm}}{\LegFnorm{\kk}{\mm}} = \kron{\nn}{\kk}.
\label{norm_cond_alf}
\end{equation}
Hough functions can thus be considered as ALFs distorted by the planet's rotation. Particularly, they converge towards the ALFs as $\spinpar $ tends to zero (non-rotating planet).

In the set $\left\{ \HoughF{\nn}{\mm}{\spinpar} \right\}$, Hough functions are indexed by the integer~$\nn$, which can be chosen as the degree of the ALF obtained in the non-rotating limit ($\spinpar = 0$). There are two types of modes \citep[e.g.][]{Unno1989}: (i) gravity modes (or g-modes), and (ii) Rossby modes (or r-modes). Gravity modes exist for $\spinpar \in \Rset$. They correspond to forced waves confined within an equatorial band. Rossby modes only exist for $\abs{\spinpar} >1$ and develop near the two poles. Unlike gravity modes, these modes never match to spherical harmonics. Therefore, it is convenient to index them with degrees $\nn<\abs{\mm}$. Hough functions are associated with the real-valued eigenvalues $\Houghval{\nn}{\mm}{\spinpar}$\mynom[S]{$\Houghval{\nn}{\mm}{\spinpar}$}{real-valued eigenfunction associated with $\HoughF{\nn}{\mm}{\spinpar}$}, which converge towards the eigenvalues of spherical harmonics as $\spinpar \rightarrow 0$ for gravity modes. Hence, $\HoughF{\nn}{\mm}{0} = \LegFnorm{\nn}{\mm}$ and $\Houghval{\nn}{\mm}{0} = \nn \left( \nn + 1 \right)$ with $\nn \geq \abs{\mm}$. As $\spinpar$ increases in absolute value, Hough functions and eigenvalues undergo significant transformations, with forced tidal waves transitioning from super-inertial to sub-inertial regimes. In practice, for any value of $\mm \in \Zset$ and $\spinpar \in \Rset$, the sets $\left\{ \HoughF{\nn}{\mm}{\spinpar} \right\}$ and $\left\{ \Houghval{\nn}{\mm}{\spinpar} \right\}$ can be numerically evaluated at once using spectral methods such as Chebyshev collocation or ALF expansion \citep[e.g.][]{Wang2016}. In this study we employ the latter approach. For illustration, \append{app:hough} presents values of the expansion coefficients and plots of Hough functions for $\mm=2$ and $\spinpar =1$.


\subsection{Vertical structure of tidal modes}

The vertical structure of the degree-$\nn$ mode associated with the Hough function $\HoughF{\nn}{\mm}{\spinpar}$ is described by \eq{pdeG_single}, in which the Laplace tidal operator given by \eq{Laplace_operator} is substituted with~$- \Houghval{\nn}{\mm}{\spinpar}$. Following \cite{LC1969}, we decompose $\Gtidev$ as 
\begin{equation}
\Gtidev \left( \xx \right) = \phiwave \left( \xx \right) \psiv \left( \xx \right),
\end{equation}
with $\phiwave$\mynom[S]{$\phiwave$}{function introduced in the change of variable leading to the wave equation}\mynom[S]{$\psiv$}{unknown of the vertical wave equation} being a background-dependent function given by 
\begin{equation}
\phiwave \left( \xx \right)  = \exp \left[ \frac{\xx}{2} - \frac{1}{2} \integ{\frac{\inumber \alphapar \dlnbet}{1 - \inumber \alphapar}  }{\xx^\prime}{0}{\xx}  \right].
\end{equation}
We recall that $\xx$ denotes the nondimensional vertical coordinate defined in \eq{vertical_coord}, and that $\alphapar$ and $\dlnbet$ -- introduced in \eq{dimensionless_numbers} -- both depend on $\xx$. Equation~(\ref{pdeG_single}) thus simplifies to 
\begin{equation}
\label{wave_eq}
\DDn{\psiv}{\xx}{2} + \kvert^2 \left( \xx \right) \psiv = \phiwave^{-1} \left( \xx \right) C \left( \xx \right),
\end{equation}
where we have introduced the normalised vertical wavenumber of the mode, $\kvert$\mynom[S]{$\kvert$}{dimensionless vertical wavenumber of the degree-$\nn$ mode}, such that
\begin{align}
\kvert^2 \left( \xx \right) = & - \frac{1}{4} \left[ \left( 1 - \frac{\inumber \alphapar \dlnbet}{1 - \inumber \alphapar}  \right)^2 + 2 \DD{}{\xx} \left( \frac{\inumber \alphapar \dlnbet }{1-\inumber \alphapar} \right) \right. \nonumber \\
\label{kvert2}
& \left. + \frac{4}{1- \inumber \alphapar} \left( \inumber \alphapar \dlnbet - \gammapar \betapar \Houghval{\nn}{\mm}{\spinpar} \right) \right], 
\end{align}
and the vertical profiles of thermal and gravitational forcings\mynom[S]{$C$}{forcing term in the wave equation},
\begin{align}
C \left( \xx \right) = & \, \frac{\betapar \Houghval{\nn}{\mm}{\spinpar} \Jtidev}{1-\inumber \alphapar}  + \frac{\alphapar \dlnbet}{1 - \inumber \alphapar}  \left( \betapar \Houghval{\nn}{\mm}{\spinpar} + \DD{}{\xx}  + \dlnbet \right) \Utidev \nonumber \\
&  - \inumber \DD{}{\xx} \left[  \left( \DD{}{\xx} + \dlnbet \right) \Utidev \right].
\end{align}
We note that $\betapar \Houghval{\nn}{\mm}{\spinpar} $ can be rewritten as 
\begin{equation}
\betapar \Houghval{\nn}{\mm}{\spinpar}  = \frac{\height}{\deptheq}, 
\end{equation}
where we retrieve the equivalent depth\mynom[S]{$\deptheq$}{equivalent depth of the degree-$\nn$ mode} used in classical theory \citep[][]{Wilkes1949,Siebert1961,LC1969},
\begin{equation}
\deptheq = \frac{\Rpla^2 \ftide^2}{\ggravi \Houghval{\nn}{\mm}{\spinpar}}.
\end{equation}

The wave equation given by \eq{wave_eq} can be solved for $\Gtidev$ by applying lower and upper boundary conditions, as discussed in \sect{sec:solution}. The vertical profiles of all tidal fields  are then straightforwardly deduced from $\Gtidev$, with the functions $\Vthetav$, $\Vphiv$, $\Vrv$, $\presstidev$, $\rhotidev$, and $\temptidev$ introduced in \eq{hough_series} being respectively expressed as
\begin{equation}
\label{Vthetav_pola}
\Vthetav = \frac{1}{\Houghval{\nn}{\mm}{\spinpar}} \left[  \DD{\Gtidev}{\xx} - \Gtidev + \inumber \left( \DD{}{\xx} + \dlnbet \right) \Utidev \right],
\end{equation}
\begin{equation}
\Vphiv = \frac{\inumber}{\Houghval{\nn}{\mm}{\spinpar}} \left[  \DD{\Gtidev}{\xx} - \Gtidev + \inumber \left( \DD{}{\xx} + \dlnbet \right) \Utidev \right],
\end{equation}
\begin{align}
\Vrv = & \frac{1}{\betapar \Houghval{\nn}{\mm}{\spinpar}} \left[ \DD{}{\xx} + \betapar \Houghval{\nn}{\mm}{\spinpar} - 1 \right] \Gtidev  \nonumber \\
& + \inumber \left[ 1 + \frac{1}{\betapar \Houghval{\nn}{\mm}{\spinpar}} \left( \DD{}{\xx} + \dlnbet \right) \right] \Utidev,
\end{align}
\begin{equation}
\presstidev = \frac{1}{\inumber \betapar \Houghval{\nn}{\mm}{\spinpar} } \left( \DD{\Gtidev}{\xx} - \Gtidev \right) \! + \! \left[  1 \! + \! \frac{1}{\betapar \Houghval{\nn}{\mm}{\spinpar} } \left( \DD{}{\xx} + \dlnbet \right) \! \right] \! \Utidev,
\end{equation}
\begin{align}
\rhotidev = & \left( 1  + \frac{\dlnbet}{1- \inumber \alphapar}  \right) \left\{   \frac{1}{\inumber \betapar \Houghval{\nn}{\mm}{\spinpar}}  \left( \DD{\Gtidev}{\xx} - \Gtidev \right) \right. \\
& \left. + \left[ 1 + \frac{1}{\betapar \Houghval{\nn}{\mm}{\spinpar}} \left( \DD{}{\xx} + \dlnbet \right)  \right] \Utidev  \right\} + \inumber \frac{\Jtidev - \gammapar \Gtidev}{1 - \inumber \alphapar} , \nonumber
\end{align}
and
\begin{align}
\label{temptidev_pola}
\temptidev = & - \frac{\dlnbet}{1-\inumber \alphapar}  \left\{ \frac{1}{\inumber \betapar \Houghval{\nn}{\mm}{\spinpar}} \left( \DD{\Gtidev}{\xx} - \Gtidev \right) \right. \\ 
& \left. + \left[ 1+ \frac{1}{\betapar \Houghval{\nn}{\mm}{\spinpar}} \left( \DD{}{\xx} + \dlnbet \right) \right] \Utidev \right\}  - \inumber  \frac{\Jtidev - \gammapar \Gtidev}{1-\inumber \alphapar}. \nonumber
\end{align}
All the equations established until now are valid in general. Particularly, any vertical profiles can be used for $\pressbg$, $\rhobg$, $\tempbg$, $\height$, and thus for the dimensionless numbers $\alphapar$, $\betapar$, $\gammapar$ and $\dlnbet$. Moreover, tidal equations include both the gravitational and thermal forcings. In the next section, we simplify them to explore thermo-gravitational atmospheric tides for the two configurations defined in \sect{ssec:physical_setup}: isentropic and isothermal.

\section{Closed-form ab initio solution}
\label{sec:solution}

\subsection{Atmospheric structure}

In general, \eq{wave_eq} can only be solved numerically because of the complex dependence of $\phiwave$, $\kvert$ and $C$ on $\xx$. Nevertheless, it is still possible to derive closed-form solutions by assuming a few simplifications, which is the purpose of this section. Such solutions have already been established in the past both for isentropic atmospheres \citep[][]{Lindzen1978,Farhat2024} and isothermal atmospheres \citep[][]{Lindzen1968,Lindzen1972,Lindzen1978,ADLM2017a,ADL2019}, but we shall demonstrate here that they actually come down to a single solution. 

The mathematical complications encountered while integrating the wave equation analytically mainly originate in the variation of $\alphapar$ with altitude. It is thus convenient to consider the cooling frequency, $\fnewton$, as globally uniform, which makes $\alphapar$ altitude-independent (\eq{dimensionless_numbers}). The variation of the vertical wavenumber with $\xx$ represents another complexity source (see \eq{kvert2}). Therefore, we require that $\kvert$ should be vertically uniform. This provides two conditions on the atmospheric structure. First the parameter $\dlnbet$ (\eq{dimensionless_numbers}) must be a constant, which implies that the pressure height is of the form 
\begin{equation}
\label{cond_1}
\height \left( \xx \right) = \heightsurf \expo{\dlnbet \xx}.
\end{equation}
In the above equation, $\dlnbet$ is a real parameter and $\heightsurf = \Rspec \Tsurf / \ggravi$ the pressure height at the planet's surface, with $\Tsurf $\mynom[S]{$\Tsurf$}{atmospheric temperature at the planet's surface} being the atmospheric temperature at the planet's surface\footnote{The atmospheric temperature at the planet's surface should not be confused with the surface temperature, as they are not generally equal. The former is the temperature of gas particles near the surface, while the second designate the temperature of the ground. The difference between the two temperatures depends on the efficiency of radiative, sensible and latent heat exchanges.}. 

Second, the product $\gammapar \betapar $ must be altitude-independent as well. Combined with \eq{cond_1}, this condition yields
\begin{equation}
\label{cond_2}
\left( \rcp + \dlnbet \right) \expo{\dlnbet \xx} = K,
\end{equation}
with $K$ being a constant. Equation~(\ref{cond_2}) clearly shows that only two values of $\dlnbet$ allow for vertically uniform wavenumbers. These values correspond to the two considered scenarios: $\dlnbet = - \rcp$ (isentropic atmosphere), and $\dlnbet= 0$ (isothermal atmosphere), the latter being the asymptotic limit of isentropic atmospheres for $\rcp \rightarrow 0$. Similarly, we note that $\gammapar =0$ if the atmosphere is isentropic and $\gammapar = \rcp$ if it is isothermal. Therefore, the parameter $\gammapar$ is no longer a free parameter but a function of $\rcp$ and $\dlnbet$ expressed as 
\begin{equation}
\gammapar = \dlnbet + \rcp.
\end{equation} 
Besides, the background density and temperature profiles are given by 
\begin{equation}
\label{background_fields_sol}
\begin{array}{ll}
\displaystyle \rhobg = \frac{\psurf}{\ggravi \heightsurf} \expo{- \left(1 + \dlnbet \right) \xx}, & \displaystyle \tempbg = \frac{\ggravi \heightsurf}{\Rspec} \expo{\dlnbet \xx}, 
\end{array}
\end{equation}
and the altitude profile by 
\begin{equation}
\label{altitude_profile}
\zz = \left\{
\begin{array}{ll}
 \displaystyle \frac{\heightsurf}{\rcp} \left( 1 - \expo{- \rcp \xx} \right) & \mbox{(isentropic)}, \\[0.3cm]
 \displaystyle \heightsurf \xx & \mbox{(isothermal)}.
\end{array}
\right.
\end{equation}

\begin{figure}[t]
   \centering
   \includegraphics[width=0.48\textwidth,trim = 0.cm 0.cm 0cm 0cm,clip]{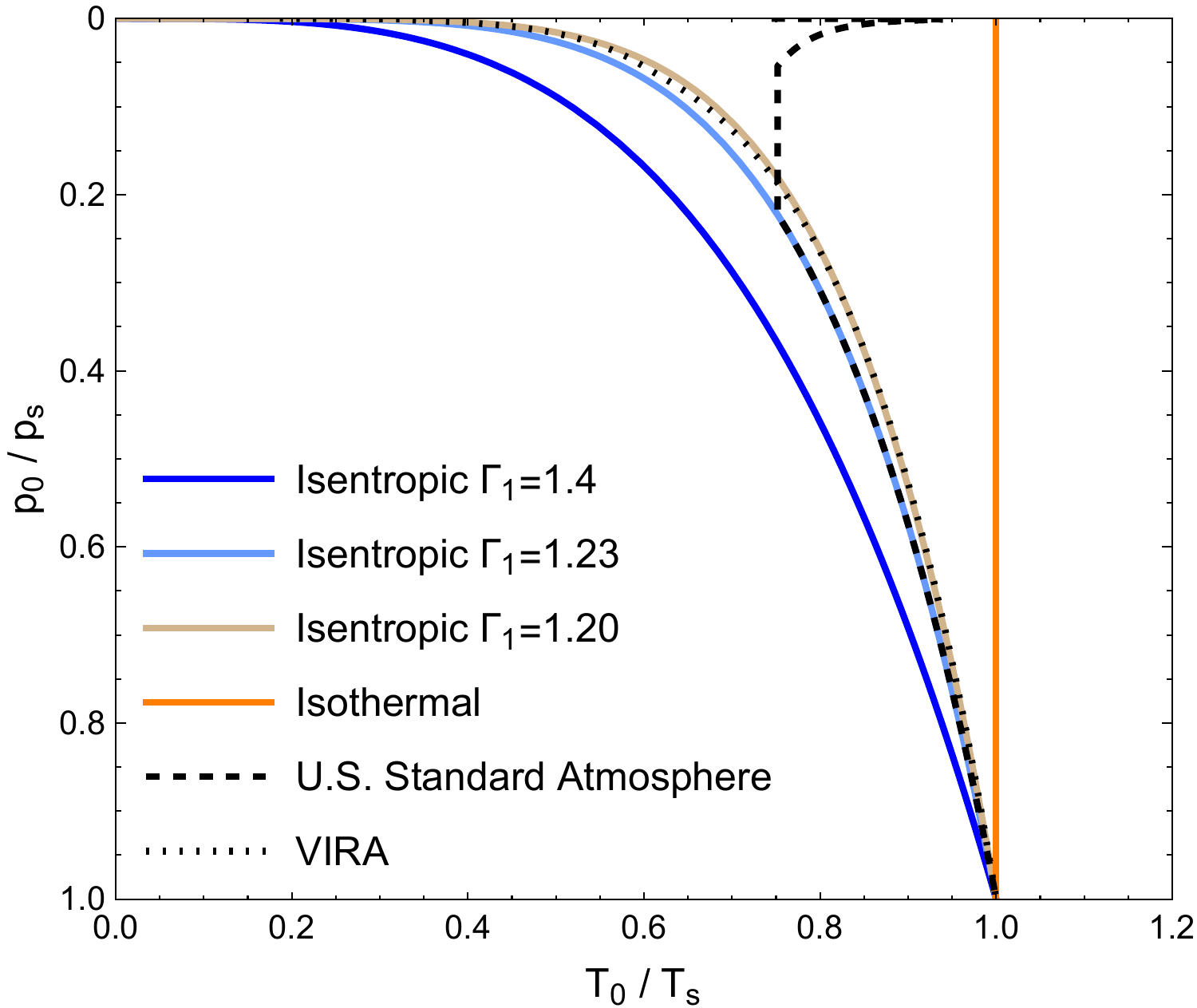}
      \caption{Vertical temperature profiles. The normalised background temperature (horizontal axis) is shown as a function of the normalised background pressure (vertical axis) for three reference atmospheres derived from \eq{background_fields_sol}: a dry-adiabatic atmosphere characterised by an ideal diatomic gas ($\adiabexp = 1.4$, solid blue line), a moist-adiabatic atmosphere with the effective adiabatic index of the Earth's troposphere ($\adiabexp = 1.23$, solid sky-blue line), an isentropic atmosphere with the effective adiabatic index of Venus' troposphere ($\adiabexp=1.2$, solid tan line) and an isothermal atmosphere (solid orange line). For comparison, the temperature profiles of the U.S Standard Atmosphere \citep[][]{USStdAtm1976} and the Venus International Reference Atmosphere \citep[e.g.][]{Seiff1985} are plotted as dashed and dotted black lines, respectively. }
       \label{fig:temp_profiles}%
\end{figure}

Figure~\ref{fig:temp_profiles} shows the vertical temperature profiles of four reference atmospheres derived from \eq{background_fields_sol}: a dry-adiabatic atmosphere characterised by an ideal diatomic gas (Isentropic $\adiabexp = 1.4$), a moist-adiabatic atmosphere with the effective adiabatic index of the Earth's troposphere (Isentropic $\adiabexp= 1.23$), an isentropic atmosphere with the effective adiabatic index of Venus' troposphere (Isentropic $\adiabexp = 1.2$), and an isothermal atmosphere (Isothermal). The value of the Earth's adiabatic index is inferred from the tropospheric lapse rate of the U.S. Standard Atmosphere 1976, $\DD{\tempbg}{\zz} = - 6.5~{\rm K \ km^{-1}} $ \citep[][]{USStdAtm1976}\footnote{\url{https://ntrs.nasa.gov/api/citations/19770009539/downloads/19770009539.pdf}}, using the formula derived for an isentropic profile in \eqs{background_fields_sol}{altitude_profile}, 
\begin{equation}
\DD{\tempbg}{\zz} = - \frac{\rcp \ggravi}{\Rspec},
\end{equation}
with $\rcp=\left( \adiabexp-1 \right)/\adiabexp$. Parameters are $\ggravi = 9.81 \ {\rm m \ s^{-1}}$ and $\Rspec = 287 \ {\rm J \ kg^{-1} \ K^{-1}}$ \citep[][]{USStdAtm1976}. The value of the adiabatic index for Venus' deep atmosphere, $\adiabexp=1.2$, is provided by Table~1-3 of \cite{Seiff1985}. The temperature profiles of the U.S. Standard Atmosphere 1976 \citep[][]{USStdAtm1976} and Venus International Reference Atmosphere \citep[VIRA; e.g.][]{Seiff1985} are also plotted for comparison. 

We remark that two atmospheres with different structures do not have the same global thermal state. All things being equal, the isothermal atmosphere holds a greater amount of heat than its isentropic counterpart. To disentangle the role played by the atmospheric structure, one needs to filter out the bias induced by differences in heat content. To do so, we introduce the mass-averaged temperature of the air column\mynom[S]{$\Tbulk$}{mass-averaged temperature of the air column},
\begin{equation}
\label{Tbulk_def}
\Tbulk = \frac{\ggravi}{\psurf} \integ{\tempbg \rhobg }{\zz}{0}{+\infty},
\end{equation}
which we call the bulk temperature in the following. The bulk temperature is proportional to the total enthalpy of the air column. It is therefore a measure of the thermal energy content of the atmosphere per unit mass. Fixing $\Tbulk$ instead of $\Tsurf$ allows the impact of atmospheric structure to be isolated from that of energy content. Using \eqs{cond_1}{background_fields_sol}, we obtain
\begin{equation}
\label{Tbulk}
\Tbulk = \frac{\Tsurf}{1- \dlnbet} = \left\{
\begin{array}{ll}
\displaystyle \frac{\Tsurf}{1+\rcp} &  \mbox{(isentropic}, \\
\Tsurf & \mbox{(isothermal)}.
\end{array}
\right.
\end{equation}


\subsection{Tide-raising potential and heating profiles}

Like the atmospheric structure, the right-hand member of \eq{wave_eq} has to be simplified. The gravitational forcing does not induce any mathematical complication. Introducing the planet-perturber distance, $d$, we remark that the tidal potential difference between the planet's surface and the top of the atmosphere scales as $\scale \height / d \ll 1$. The tidal potential can thus be taken uniform across the vertical direction. It follows
\begin{equation}
\begin{array}{lcl}
\displaystyle \Utidev = \frac{\Usurf}{\ggravi \height}, & \mbox{and} & \displaystyle \left( \DD{}{\xx} + \dlnbet \right) \Utidev \approx 0,
\end{array} 
\end{equation}
with $\Usurf$\mynom[S]{$\Usurf$}{degree-$\nn$ tide-raising potential at planet's surface} denoting the degree-$\nn$ Hough component of the tide-raising potential evaluated at planet's surface. As discussed by \cite{LC1969}, the vertical profile of the thermal forcing term can be much more complex than that of the gravitational tidal potential because it is determined by the local absorption properties of the atmospheric mixture. Following earlier studies \citep[][]{ADLM2017a,ADL2019,Farhat2024}, we consider an exponentially decaying profile of the thermotidal heating per unit mass per unit time,
\begin{equation}
\label{Jtidev_function}
\Jtidev = \frac{\rcp \Jsurf}{\ftide \ggravi \height} \expo{- \bj \xx},
\end{equation}
where $\Jsurf$\mynom[S]{$\Jsurf$}{degree-$\nn$ heating at the planet's surface} is the heating evaluated at planet's surface and $\bj$\mynom[S]{$\bj$}{vertical variation rate of the tidal heating} its vertical variation rate. 

We note that $\Jsurf$ implicitly depends on $\bj$, as $\bj$ characterises the absorbed heat distribution over the air column. The relationship between the two parameters is straightforward: assuming that a fraction $\opacity$ of the total stellar flux impinging the atmosphere is absorbed, we write
\begin{equation}
\label{relation_Js_deltaF}
 \integ{\Jsurf \expo{-\bj \xx} \rhobg \height}{\xx}{0}{+ \infty} = \opacity  \fluxtiden ,
\end{equation}
where $\fluxtiden$\mynom[S]{$\fluxtiden$}{degree-$\nn$ Hough mode of the incident stellar flux} represents the degree-$\nn$ Hough mode of the incident flux.
The heating at planet's surface, $\Jsurf$, is thus expressed as
\begin{equation}
\label{Jsurf}
\Jsurf = \frac{\ggravi \left( 1 + \bj \right) \opacity \fluxtiden}{\psurf}. 
\end{equation}
The functional form of $\Jtidev$ defined by \eq{Jtidev_function} allows for studying a broad range of heating profiles. With $\bj = 0$, the tidal heating is uniformly distributed over the air column. With $\bj = + \infty$, it is concentrated at the planet's surface. The latter case typically corresponds to the Dirac distribution assumed by \cite{DI1980} to model Venus' thermal tides. In general, the coefficients $\fluxtiden$ introduced in \eq{relation_Js_deltaF} depend on the system's frequencies and orbital parameters (mean motion, eccentricity, obliquity, semi-major axis, etc.). They are evaluated in \papertwo for a zero-obliquity planet orbiting its host star circularly.

We further lighten expressions by introducing the dimensionless quantities $\betasurf = \betapar_{\xx=0}$\mynom[S]{$\betasurf$}{$\betapar$ at the planet's surface} and $\lambdapar$\mynom[S]{$\lambdapar$}{dimensionless complex coefficient parametrising $\phiwave$}, such that $\phiwave = \expo{\lambdapar \xx}$, defined as
\begin{align}
& \betasurf = \frac{\ggravi \heightsurf}{\Rpla^2 \ftide^2}, & \lambdapar = \frac{1 - \inumber \alphapar \left( 1 + \dlnbet \right)}{2  \left(1-\inumber \alphapar \right)}.
\end{align}
The squared vertical wavenumber, given by \eq{kvert2}, is thus rewritten as
\begin{equation}
\label{kvert2_solution}
\kvert^2 =   \frac{\gammapar \betapar \Houghval{\nn}{\mm}{\spinpar} - \inumber \alphapar \dlnbet }{1 - \inumber \alphapar} - \lambdapar^2.
\end{equation}
Since $\kvert$ is defined up to a minus sign, we take it such that $\Im \left\{  \kvert \right\} \geq 0$. Besides, it is noteworthy that $\Im \left\{ \kvert \right\} \neq 0$ as far as $\fnewton \neq 0$, namely in the presence of dissipation. 

\subsection{Boundary conditions}
\label{ssec:boundary_conditions}
The solution of the homogeneous equation associated with \eq{wave_eq} (i.e. such that $C=0$) is expressed as 
\begin{equation}
\psiv \left( \xx \right) = \Aconst \expo{\inumber \kvert \xx} + \Bconst \expo{-\inumber \kvert \xx}, 
\end{equation}
where $\Aconst$ and $\Bconst$ denote two complex-valued integration constants. Two boundary conditions are thus necessary to solve \eq{wave_eq}. For the upper boundary condition, one requires that the kinetic energy density, $\ekin = \frac{1}{2} \rhobg \Vvect^2 $\mynom[S]{$\ekin$}{kinetic energy density}, shall remain bounded at $\xx \rightarrow + \infty$ \citep[][]{Wilkes1949,Siebert1961,LC1969}. Since the only diverging term of $\Vvect$ scales as $\scale \exp  \left\{ \left[\Im \left( \kvert \right) + \Re \left( \lambdapar \right) \right] \xx \right\}$ and the background density as $\scale \exp \left[ - \left( 1 + \dlnbet \right) \xx \right]$, the dominating term of $\ekin$ at $\xx \rightarrow + \infty$ scales as
\begin{equation}
\ekin \scale \exp \left\{ \left[ 2 \Im \left( \kvert \right) + 2 \Re \left( \lambdapar \right) - \left( 1 + \dlnbet \right) \right] \xx \right\}.
\end{equation}
We note that $2 \Re \left( \lambdapar \right) - \left( 1 + \dlnbet \right) = - \dlnbet / \left( 1 + \alphapar^2 \right)  \geq 0$. Moreover, $\Im \left( \kvert \right) > 0$ by definition. It follows 
\begin{equation}
2 \Im \left( \kvert \right) + 2 \Re \left( \lambdapar \right) - \left( 1 + \dlnbet \right)>0,
\end{equation} 
which implies $\ekin  \rightarrow + \infty$ at $\xx \rightarrow + \infty$. As a consequence, the integration constant in factor of the diverging term of $\psiv$ has to be set to zero ($\Bconst=0$) to make the kinetic energy density remain bounded at the upper boundary. We emphasise that this boundary condition can be used only if $\Im \left\{ \kvert \right\} \neq 0$, which is satisfied as far as $\alphapar \neq 0$. When adiabatic compression is assumed ($\alphapar =0$), $\kvert^2$ is real. In this scenario, the above boundary condition can be used if $\kvert^2 <0$ (evanescent regime) solely. If $\kvert^2>0$ (oscillatory regime), the kinetic energy density is already bounded at $\xx \rightarrow + \infty$ since the atmosphere behaves as a non-damped harmonic oscillator. To treat this specific case, one can apply a radiation condition at the upper boundary, which reflects the fact that tidal waves only carry energy upwards \citep[e.g.][]{Wilkes1949}. 

To compute the second integration constant, $\Aconst$, one simply requires that fluid particles do not cross the planet's surface, meaning that $\Vrv = 0$ at $\xx = 0$. This leads to 
\begin{equation}
\label{Aconst_solution}
\Aconst \! = \! \frac{\left( \bj+1-\betasurf \Houghval{\nn}{\mm}{\spinpar} \right) \psitherm + \left( 1 - \betasurf \Houghval{\nn}{\mm}{\spinpar} \right) \psigrav - \inumber \frac{\betasurf \Houghval{\nn}{\mm}{\spinpar}}{\ggravi \heightsurf}  \Usurf}{\inumber \kvert + \lambdapar - 1 + \betasurf \Houghval{\nn}{\mm}{\spinpar}}  \! ,
\end{equation}
where $\psitherm$\mynom[S]{$\psitherm$}{constant factor associated with tidal heating} and $\psigrav$\mynom[S]{$\psigrav$}{constant factor associated with tidal gravitational forcing} are constant factors accounting for tidal heating and gravitational forcing, respectively. These factors are expressed as 
\begin{align}
\psitherm & = \frac{ \Lambn \rcp \Jsurf}{\ftide^3 \Rpla^2 \left(1 - \inumber \alphapar \right) \left[ \left( \lambdapar + \bj \right)^2 + \kvert^2 \right]}, \\
 \psigrav & = \frac{\Lambn\alphapar \dlnbet \Usurf}{\ftide^2 \Rpla^2 \left(1 - \inumber \alphapar \right) \left( \lambdapar^2 + \kvert^2\right) }.
\end{align}

\subsection{Vertical distributions of tidal fields}
The various approximations and modelling choices described above yield analytical expressions for $\psiv$ and $\Gtidev$, which are given by
\begin{equation}
\psiv = \Aconst \expo{\inumber \kvert \xx} + \psitherm \expo{- \left( \lambdapar + \bj \right) \xx} + \psigrav \expo{- \lambdapar \xx},
\end{equation}
and
\begin{equation}
\label{solution_Gtidev}
\Gtidev = \Aconst \expo{\left(  \inumber \kvert + \lambdapar \right) \xx} + \psitherm \expo{-  \bj \xx} + \psigrav. 
\end{equation}
The vertical distributions of all the tidal fields are then straightforwardly obtained by substituting $\Gtidev$ with \eq{solution_Gtidev} in \eqsto{Vthetav_pola}{temptidev_pola}. We end up with
\begin{align}
\label{Vthetav_solution}
\Vthetav =   \frac{1}{\Lambn} & \left[  \Aconst \left( \inumber \kvert + \lambdapar -1 \right) \expo{\left(\inumber \kvert + \lambdapar \right) \xx } \right. \nonumber \\
& \left. - \left( \bj + 1 \right) \psitherm \expo{-\bj \xx} - \psigrav \right],
\end{align}
\begin{align}
\Vphiv = \frac{\inumber}{\Lambn}  & \left[ \Aconst \left( \inumber \kvert + \lambdapar - 1 \right) \expo{\left( \inumber \kvert + \lambdapar \right) \xx}  \right. \nonumber \\
& \left. - \left( \bj + 1 \right) \psitherm \expo{- \bj \xx} - \psigrav  \right],
\end{align}
\begin{align}
\Vrn = & \frac{1}{\betapar \Lambn}  \left[  \Aconst \left( \inumber \kvert + \lambdapar -1 \right) \expo{\left( \inumber \kvert + \lambdapar \right) \xx} \right. \\
& \left. - \left( \bj \! + \! 1 \! - \! \betapar \Lambn \right) \psitherm \expo{- \bj \xx} - \left( 1\! - \! \betapar \Lambn \right) \psigrav  \right] + \inumber \frac{\Usurf}{\ggravi \height}, \nonumber 
\end{align}
\begin{align}
\label{presstidev}
\presstidev =  \frac{1}{\inumber \betapar \Lambn} & \left[   \Aconst \left( \inumber \kvert + \lambdapar - 1 \right) \expo{\left( \inumber \kvert + \lambdapar \right) \xx} \right. \\
& \left. - \left( \bj + 1 \right) \psitherm \expo{- \bj \xx} - \psigrav   \right] + \frac{\Usurf}{\ggravi \height}, \nonumber 
\end{align}
\begin{align}
\rhotidev = & \frac{1}{\inumber \betapar \Lambn \left( 1 - \inumber \alphapar \right)} \left\{ \frac{1}{\ggravi \height} \left[ \left( 1 + \dlnbet - \inumber \alphapar \right) \Usurf  - \frac{\betapar \Lambn \rcp \Jsurf}{\ftide}  \right] \right. \nonumber \\
& + \Aconst \left[ \left( 1 + \dlnbet - \inumber \alphapar \right) \left( \inumber \kvert + \lambdapar - 1 \right) + \gammapar \betapar \Lambn \right] \expo{\left( \inumber \kvert + \lambdapar \right) \xx}   \nonumber \\
& - \psitherm \left[ \left( \bj+1 \right) \left( 1 + \dlnbet - \inumber \alphapar \right) - \gammapar \betapar \Lambn \right] \expo{- \bj \xx} \nonumber \\
&\left.  - \psigrav \left[ 1 + \dlnbet - \inumber \alphapar - \gammapar \betapar \Lambn \right] \right\} , 
\end{align}
and
\begin{align}
\label{temptidev_solution}
\temptidev = & - \frac{1}{\inumber \betapar \Lambn \left( 1 - \inumber \alphapar \right)} \left\{ \frac{\inumber \betapar \Lambn}{\ggravi \height} \left( \dlnbet \Usurf + \inumber \frac{\rcp \Jsurf}{\ftide}  \right)  \right.  \\
&+ \Aconst \left[ \dlnbet \left( \inumber \kvert + \lambdapar - 1 \right) + \gammapar \betapar \Lambn \right] \expo{\left( \inumber \kvert + \lambdapar \right) \xx}    \nonumber \\
& \left.  - \psitherm \left[  \dlnbet \left( \bj \! + \! 1 \right) - \gammapar \betapar \Lambn \right] \expo{- \bj \xx} - \psigrav \left( \dlnbet \! - \! \gammapar \betapar \Lambn \right) \right\} \! . \nonumber 
\end{align}


\subsection{Resonances of tidally excited Lamb waves}
\label{ssec:resonances_lamb}
The obtained closed-form solution (\eqsto{solution_Gtidev}{temptidev_solution}) describes the tidally forced oscillations of pressure, density, temperature, and velocities near a state of equilibrium. These oscillations are restored by compressibility and Coriolis forces in general, and complemented by Archimedean forces in the isothermal scenario, where the atmosphere is stably stratified. The atmospheric tidal response thus takes the form of a series of forced Lamb waves, namely horizontally propagating acoustic waves of planetary-scale wavelengths \citep[][]{Bretherton1969,Lindzen1972}. Analogous to surface gravity waves propagating in Earth's oceans, Lamb waves can be resonantly excited by tidal forces. This occurs when the tidal frequency equalises the eigenfrequency of a tidal mode. 

Considering the expressions obtained for the vertical distributions of tidal fields, given by \eqsto{Vthetav_solution}{temptidev_solution}, we remark that the solution's component associated with wave propagation is the term varying as $\exp \left[ \left( \inumber \kvert + \lambdapar \right) \xx \right] $. The amplitude of this term is modulated by the factor $\Aconst$, given by \eq{Aconst_solution}. Resonance thus occurs for values of tidal frequencies minimising the denominator $D \left( \ftide \right)$ of $\Aconst$ in absolute value,
\begin{equation}
D \left( \ftide \right) = \abs{\inumber \kvert + \lambdapar - 1 + \betasurf \Houghval{\nn}{\mm}{\spinpar} }.
\end{equation}
Since resonant excitation does not require energy dissipation, it is convenient to ignore dissipative processes by placing ourselves in the limit of adiabatic compression ($\alphapar \rightarrow 0$)\footnote{This assumption might seem contradictory with the discussion of \sect{ssec:boundary_conditions}, in which we stress that the used upper boundary condition holds only if $\alphapar \neq 0$. Nevertheless, we remark that dissipative processes act as small perturbations for the resonance frequency as far as $\abs{\alphapar} \ll 1 $. Thus, as $\alphapar \rightarrow 0$, the minima of $D$ asymptotically converge towards those calculated in the absence of dissipation, meaning that the derivations of \sect{ssec:resonances_lamb} hold.}. The squared modulus of the denominator thus simplifies to 
\begin{equation}
\label{squared_denominator}
D^2 \left( \ftide \right) = \abs{\kvert}^2 + \left( \betasurf \Lambn - \frac{1}{2} \right)^2 - 2 \left( \betasurf \Lambn - \frac{1}{2} \right) \Im \left( \kvert \right). 
\end{equation}
In the absence of dissipation, the squared vertical wavenumber given by \eq{kvert2_solution} is real and reduces to
\begin{equation}
\label{kvert2_reson}
\kvert^2 = \gammapar \betasurf \Lambn - \frac{1}{4}. 
\end{equation}
It is worth noting that we have replaced $\gammapar \betapar$ by $\gammapar \betasurf$ in \eq{kvert2_reson}. This results from the specific atmospheric structures described by our closed-form solution. In the isentropic case, $\gammapar=0$; in the isothermal one, $\gammapar=\rcp$ and $\betapar=\betasurf$. As a consequence, $\gammapar \betapar = \gammapar \betasurf$ in both cases. 

Notwithstanding the turning point ($\kvert=0$), we must examine two cases: (i) $\kvert^2 > 0$ and (ii) $\kvert^2 <0$. In the first case ($\kvert^2>0$), substituting $\abs{\kvert}^2 = \kvert^2$ and $\Im \left( \kvert \right) = 0$ in \eq{squared_denominator}, we obtain that $D^2 = 0$ for $\betasurf = \left( 1 - \gammapar \right) / \Lambn$. We recover this condition in the second case ($\kvert^2 < 0$) by substituting $\abs{\kvert}^2 = - \kvert^2$ and $\Im \left( \kvert \right) = \sqrt{ - \kvert^2}$ in \eq{squared_denominator}. As a consequence, $\Aconst$ is singular for values of tidal frequencies such that
\begin{equation}
\label{condition_resonance_ftide}
\ftide^2 = \frac{\ggravi \heightsurf \Lambn}{\Rpla^2 \left( 1 - \gammapar \right)}. 
\end{equation}
We recover here the generalised form of the condition obtained by Lamb in the non-rotating case \citep[][§314, Eq.~(3), p.~557 and §315, Eq.~(5), p.~558]{Lamb1932book}, which is also given by Eq.~(4.14-1.15) of \cite{Siebert1961}. The full derivation of \eq{condition_resonance_ftide} is provided in \append{app:calculation_resonance_freq}.

Since $0 \leq \gammapar < 1$, resonances exist only if $\Lambn >0$. For negative $\Lambn$, the singularity vanishes. Also, we emphasise that the definition of the resonance frequency given by \eq{condition_resonance_ftide} is implicit in general as the eigenvalues associated with tidal modes depend on $\ftide$ through $\spinpar$. However, $\spinpar$ ceases to vary in the asymptotic regime of rapid rotation, where $\spinrate$ is much greater than the other frequencies characterising the orbital dynamics of the planet-perturber system. In this regime, $\ftide \approx \qint \spinrate$ and $\spinpar \approx 2 / \qint$, with $\qint$ being a integer, which leads to frequency-independent $\Lambn$. Therefore, the resonance frequency of the degree-$\nn$ tidally excited Lamb wave is expressed a\mynom[S]{$\freson$}{resonance frequency of the degree-$\nn$ forced Lamb wave}
\begin{equation}
\label{freson_lamb}
\freson = \frac{1}{\Rpla} \sqrt{\frac{\ggravi \heightsurf \abs{\Lambn}}{1 - \gammapar}},
\end{equation}
which is complemented with the sign of the eigenvalue\mynom[S]{$\Sn$}{sign of $\Lambn$},
\begin{equation}
\Sn = \sign \left( \Lambn \right).
\end{equation}
The resonant regime corresponds to $\Sn = + 1$, with the resonance occurring for $\ftide = \freson$; and the non-resonant regime to $\Sn = -1$. 

Because of adiabatic compression ($\alphapar=0$), the energy is stored by the degree-$\nn$ mode without being dissipated at $\ftide = \freson$ in the resonant regime. As a consequence, tidal fields are all infinite at the resonance. The damping effect of dissipative processes eliminates singularities and regularises the atmosphere's tidal response by capping the resonantly excited tidal fields at finite amplitudes. Considering \eq{freson_lamb}, we remark that the resonance frequencies in the isothermal and isentropic configurations, denoted by $\freson^{\rm iso}$ and $\freson^{\rm isen}$, respectively, differ by a factor of
\begin{equation}
\frac{\freson^{\rm iso}}{\freson^{\rm isen}} = \sqrt{\adiabexp}.
\end{equation}
For equal bulk temperatures (\eq{Tbulk_def}), this ratio becomes
\begin{equation}
\frac{\freson^{\rm iso}}{\freson^{\rm isen}} = \sqrt{\frac{\adiabexp}{1 + \rcp}}.
\end{equation}


\section{Model limitations}
\label{sec:model_limitations}

Our closed-form solution for the atmospheric tidal response is derived at the cost of several significant simplifications. Below, we discuss the main limitations of the model that arise from these assumptions.


\paragraph{{\large \it General circulation and tide-mean flow interactions.}} The most obvious simplification is the absence of large scale circulation in the used framework. The whole planet, including the atmosphere, is assumed to rotate as a solid body. Winds and jets are ignored. This simplification is justified by the net separation of time and spatial scales between the tidal and mean flows. The formers take the form of planetary scale waves with typical periods ranging from days to months, while the latter are associated with smaller spatial scales and larger evolution times. This separation of scales tends to decouple tides from large-scale circulation. However, energy and angular momentum transfers may become important if a wide atmospheric region is rotating differentially with respect to the solid part, as described by Eliassen-Palm fluxes \citep[][Chapter~10]{Vallis2017}. Tide-mean flow interactions have been investigated with GCMs \citep[e.g.][]{Fesen1993,Miyahara1993,MM1997,HR2001,Grieger2002}, which are required to compute the large scale circulation and tidal waves self-consistently. However, numerous intermediate models of various complexities have also been developed since the 1970s to investigate the way mean flows affect Solar and Lunar tides on Earth. In these models, tidal equations are integrated numerically with spatially dependent mean winds, temperature and dissipative processes \citep[][]{LH1974,FG1979,Forbes1982,Vial1986,Hagan1995,Hagan1999,Hagan2001,HF2002,WA1997,Huang2007,Reddimalla2025}. On Earth, theoretical studies suggest that mean flows essentially influence tidal wind oscillations at high altitudes (30-100~km) while leaving their structure qualitatively unchanged \citep[e.g.][]{LH1974,Reddimalla2025}.

\paragraph{{\large \it Horizontal variations of the atmospheric structure.}} The background fields (pressure, temperature, density, pressure height) are assumed to vary across the vertical direction solely. Dependences upon horizontal coordinates are neglected. This configuration describes well thick atmospheres with efficient horizontal heat transport. However, thin atmospheres are more sensitive to local energy inputs. Typically, the Earth atmospheric structure varies latitudinally due to differential insolation between the equator and the poles \citep[e.g.][Sect.~9.2]{Pierrehumbert2010}. This horizontal variation of background fields may alter the large-scale structure of the atmospheric tidal response, thus violating the spherical symmetry assumed in the classical theory. That said, calculations performed for atmospheric tides on Earth using sophisticated models show little sensitivity to temperature details \citep[e.g.][]{LH1974}.



\paragraph{{\large \it Friction of tidal flows against the surface.}} A further limitation of the theory concerns the handling of friction of tidal flows against the ground, which notably encompasses the energy losses due to internal gravity waves generated by mountains \citep[e.g.][]{Navarro2018}. Whereas noteworthy contributions have been made by many authors to enrich numerical methods with mechanisms acting on atmospheric tides \citep[background winds, horizontal temperature gradients, composition, hydromagnetic coupling, radiative cooling, eddy and molecular diffusion, ion drag; e.g.]{LH1974,Forbes1982,Vial1986,WA1997,Hagan1995}, frictional forces are commonly ignored in analytical solutions for mathematical reasons. Typically, surface friction is more complex to incorporate in analytical models than Newtonian cooling, which is the reason why the latter is often employed as an effective term to mimic dissipative energy losses in general \citep[e.g. in][]{ZW1987}. Yet, such interactions are implicitly required in order to dynamically couple the atmosphere to the planet's solid regions. Otherwise, the lowest layers of a free atmosphere would be accelerated in differential rotation, thus breaking the solid rotation approximation, which is not observed on Earth and Venus. Despite its complexity, the influence of surface friction on tidal flows can be investigated analytically to a certain extent by inserting a Rayleigh drag into the horizontal momentum equation, given by \eqs{momentum_theta}{momentum_phi}, following \cite{Volland1974a} and \cite{Auclair2019}. With this additional term, the real-valued spin parameter defining the Laplace operator (\eq{dimensionless_numbers}) becomes a complex number, which leads to complex-valued Hough functions and eigenvalues. This allows for demonstrating that friction regularises the tidal response  by acting to annihilate the distortion effect of Coriolis acceleration. However, complex Hough functions  still remain poorly understood to our knowledge, contrary to the standard real ones, which have long been proved to form a complete set of orthogonal functions \citep[][]{Holl1970}. They shall therefore be used with caution.

\paragraph{{\large \it Internal radiative exchanges and heating profiles.}} Radiative transfer plays a key role in the tidal response of planetary atmospheres. Unlike friction, it can be partly included in analytical solutions within the classical tidal theory by means of Newtonian cooling. However, the Newtonian cooling coefficient ($\fnewton$) has to remain simply defined as a function of spatial coordinates, otherwise the tidal equations given by \eqsto{momentum_theta}{ideal_gas_equation} cannot be solved analytically. The globally uniform coefficient that we adopted to derive our solution provides a crude description of radiative losses. Particularly, it does not account for the complex radiative coupling of atmospheric layers: each layer partly scatters and absorbs fluxes emitted by other layers, while it radiatively releases energy downwards and upwards in turn. Absorption and scattering both depend on the wavelength of light fluxes and on the local chemical properties of the gas mixture \citep[e.g.][]{Heng2017book}. Calculating radiative exchanges self-consistently thus requires to adopt numerical approaches, such as GCMs. That said, the treatment of radiative transfer in GCMs is also commonly based on approximations, such as the correlated-$k$ distribution method for instance \citep[][]{LO1991}. When internal radiative exchanges are not modelled, realistic heat profiles can still be introduced in tidal equations using the prescriptions provided by radiative transfer models \citep[e.g.][]{VR2013}.


\paragraph{{\large \it Soil-atmosphere heat exchanges.}} Analogous to radiative transfer, soil-atmosphere heat exchanges are not incorporated in analytical models as they preclude closed-form solutions in general. This is the case of the present study, where direct absorption of the incident stellar flux solely is considered. Yet, tides may be influenced by the radiative, sensible, and diffusive heat exchanges taking place between the ground and atmospheric layers. For example, using a simplified soil model without atmospheric feedback, we showed in previous works that soil diffusion might delay the indirect thermal forcing generated by the planet's surface, thus altering the frequency behaviour of the associated thermotidal torque \citep[][]{ADL2019,Farhat2024}. 
In addition, phase changes can have a major impact on atmospheric tides. Typically, \cite{Lindzen1978} hypothesised that latent heat releases (evaporation, rainfalls) might represent a significant excitation mechanism for the Earth's semidiurnal tidal oscillation, which later received support from global analyses of meteorological records \citep[][]{Hamilton1981,Palumbo1998} and GCM simulations \citep[][]{DG2024}. These effects will be further explored in \papertwo and \paperthree. 

\paragraph{{\large \it Non-linear mechanisms.}} Last but not least, nonlinearities are formally excluded from the classical tidal theory. Yet, most key mechanisms influencing thermal tides are fundamentally non-linear. Typically, the intensity of thermal radiation scales as $\scale \temp^4$, while forces associated with turbulent friction of tidal flows against the planet's surface are complex functions of velocities \citep[e.g.][]{HB1993}. Several authors investigated the impact of non-linearity on the Earth's diurnal and semidiural tides, either by complementing the linear tidal equations with non-linear terms \citep[][]{Dickinson1969,Huang2007} or by using GCMs \citep[e.g.][]{Kahler1989,Fesen1993,Miyahara1993,MM1997,Grieger2002}. They showed that non-linearity can be responsible for discrepancies between observations and linear solutions, by significantly altering the background winds and mean temperature in the Earth's mesosphere for instance \citep[][]{Huang2007}. However, little attention has been given so far to the impact of non-linearity in terms of planetary-scale mass redistribution, although the latter determines how atmospheric tides influence the rotation of rocky planets. Notwithstanding extreme tidal heating or gravitational forces, tides behave as a small disturbance in the vicinity of a state of equilibrium in terms of global mass redistribution, which is consistent with the linear approximation. However, this approximation may no longer holds when Lamb waves are resonantly excited (see \sect{ssec:resonances_lamb}), given that the associated resonant amplification might theoretically increase the amplitude of tidal oscillations by several orders of magnitude. As soon as tidal fields become comparable with background fields in order of magnitude, they are strongly damped by non-linear dissipative mechanisms. As a result, the resonant tidal amplification should theoretically not be able to reach the arbitrarily high values predicted by linear theory as $\alphapar \rightarrow 0$. Instead, one should only consider the threshold attained in the linear approximation as an upper bound of the actual resonant amplitude. 
The few numerical solutions obtained so far from GCM simulations tend indirectly to corroborate this statement since none of them shows strong amplification when the predominant tidal mode is resonantly excited \citep[][]{ADL2019,Wu2023,DG2024}.



\section{Conclusions}
\label{sec:conclusions}

In this study, we have investigated analytically the response of planetary atmospheres to tidal stellar heating and gravitational forces in the framework of linear theory. The adopted method was aimed to couple tides with orbital dynamics in planetary evolution models, which requires computationally efficient modelling approaches. As a first step, we generalised the classical tidal theory to dissipative cases using Newtonian cooling to model energy dissipation. In this framework, each tidal field is expanded as a series of tidal modes written as functions of separated variables. The latitudinal components of these modes, called Hough functions, are the solutions of the eigenvalues-eigenvectors problem defined by Laplace's tidal equation with regularity boundary conditions at the poles. The associated eigenvalues are real-valued numbers that depend on both the tidal frequency and the planetary spin rate. We thus established the equations governing the tidal dynamics and those relating tidal fields to each other. These equations were then used to characterise the horizontal and vertical structures of tidal modes. 

As a second step, we showed that several analytical tidal models used in literature actually come down to one single model, which we detailed explicitly (\eqsto{solution_Gtidev}{temptidev_solution}). Mathematical constraints restrict this model to two specific atmospheric strutures -- those long known, since the foundations of classical theory, to permit analytic treatment.
In the first configuration, the atmosphere is strictly isentropic, meaning that the air column is unstratified from bottom to top, while the second configuration corresponds to isothermal atmospheres, which are associated with stable stratification. These two model atmospheres frame the typical temperature profiles of rocky planets, thus allowing the influence of the atmospheric structure on the tidal response to be quantified despite its complexity. In both configurations, tidal modes associated with positive eigenvalues can be resonantly excited as they are dominated by large scale compressibility waves, called Lamb waves, which are analogous to long-wavelength surface gravity waves in oceans. By analysing the frequency behaviour of our model, we recovered the expressions for the resonance frequencies derived in earlier studies \citep[e.g.][]{Siebert1961,Lamb1932book} as functions of the relevant physical parameters (\eq{freson_lamb}).

The model limitations originate in the simplifications assumed to solve tidal equations analytically. Mean flows and their coupling to tidal waves are ignored, as well of horizontal variations of the atmospheric structure. Dissipative mechanisms are incorporated in a simplified way, using Newtonian cooling, which does not account for the complexity of surface friction, internal radiative transfer or soil-atmosphere heat exchanges. Also, the thermal forcing is only due to the absorption of the incident stellar flux in the provided closed-form solution of thermal tides. Despite these simplifications, the model captures key features of the atmospheric tidal response -- such as the aforementioned Lamb wave resonance -- and can therefore be used as a versatile tool to probe the parameter space. This approach complements the numerical solutions derived from GCM simulations, which are more exhaustive but computationally demanding. 

In \papertwo, the theory will be used to express analytically the tidal torque induced by semidiurnal thermal tides on rocky planets as a function of the tidal frequency and the parameters of the physical setup (planet radius, surface gravity, specific gas constant, atmospheric temperature at the planet's surface, adiabatic index). Particularly, by applying the analytical model to the Earth and Venus, we will show that it agrees well with GCM solutions elaborated in previous works for these two planets, as it robustly captures the frequency dependence of the torque both in the low and high-frequency ranges. Beyond the Earth and Venus, the analytical approach presented here is highly relevant to studies dealing with atmospheric tides on extrasolar rocky planets because it provides a concise framework to examine the tidal dynamics of thin planetary atmospheres as well as the associated long-term thermo-orbital evolution.



\begin{acknowledgements}
M.~Farhat is supported by the Miller Institute for Basic Research in Science at UC Berkeley through a Miller Research Fellowship. J.~Laskar acknowledges funding from the European Research Council (ERC) under the European Union's Horizon 2020 research and innovation programme (Advanced Grant AstroGeo-885250). This research has made use of NASA's Astrophysics Data System. 
\end{acknowledgements}

\bibliographystyle{aa}  
\bibliography{references} 

\clearpage

\appendix

\section{Nomenclature}
\label{app:nomenclature}

The notations introduced in the main text are listed below in order of appearance.

\vspace{-1.2cm}

\printnomenclature

\section{Hough functions and eigenvalues}
\label{app:hough}

\def\hraisebox{2.5cm}
\def\wpanel{0.3\textwidth}
\def\hpanel{0.26\textwidth}
\begin{figure*}[t]
   \centering
   \raisebox{\hraisebox}[1cm][0pt]{\rotatebox[origin=c]{90}{$\HoughF{\nn}{2}{1} \left( \mu \right)$}}
   \includegraphics[height=\hpanel,trim = 0.9cm 0.cm 3cm 0.3cm,clip]{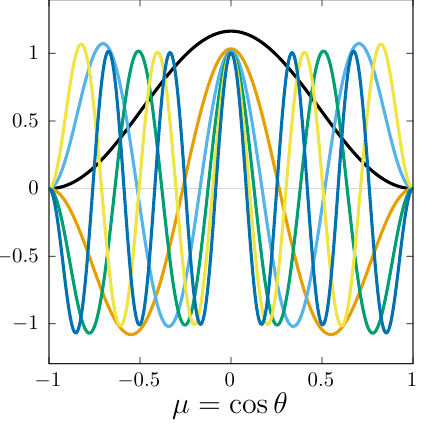}
   \raisebox{\hraisebox}[1cm][0pt]{\rotatebox[origin=c]{90}{$\Houghcol{\nn}{2}{1} \left( \mu \right)$}}
   \includegraphics[height=\hpanel,trim = 1.03cm 0.cm 3cm 0.3cm,clip]{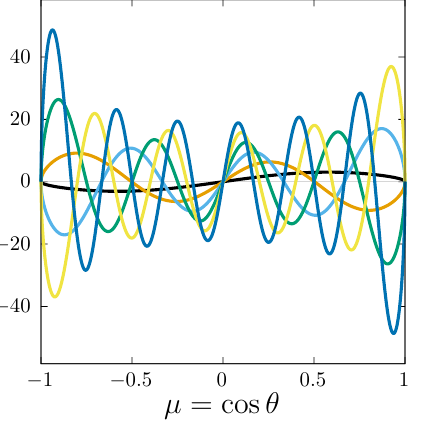}
   \raisebox{\hraisebox}[1cm][0pt]{\rotatebox[origin=c]{90}{$\Houghlon{\nn}{2}{1} \left( \mu \right)$}}
   \includegraphics[height=\hpanel,trim = 1.03cm 0.cm 3cm 0.3cm,clip]{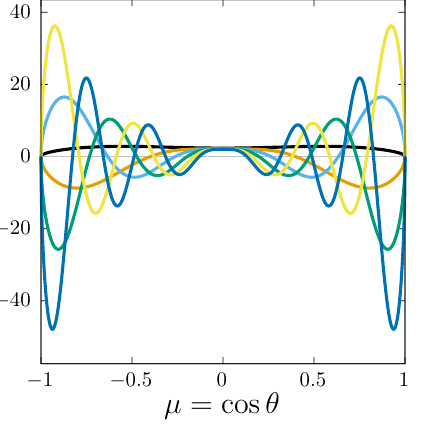}
   \includegraphics[height=\hpanel,trim = 8.7cm 0.cm 0cm 0.3cm,clip]{auclair-desrotour_figB1c}
      \caption{Symmetric Hough functions and the associated velocity-field functions for $\mm=2$, $\spinpar=1$, and degrees ranging between $2$ and $12$. The plotted functions, $\HoughF{\nn}{2}{1} \left( \mu \right)$, $\Houghcol{\nn}{2}{1} \left( \mu \right)$ and $\Houghlon{\nn}{2}{1} \left( \mu \right)$, are defined by \eqs{hough_velocity}{equation_laplace}. They were evaluated using the spectral method detailed in \cite{Wang2016}.}
       \label{fig:hough_functions}%
\end{figure*}

\begin{table*}[h] 
\centering 
\begin{small}
 \caption{\label{tab:table_coefs} Expansion coefficients connecting the normalised Hough functions, $\Theta_{n}^{m,\nu}$, with the normalised ALFs, $P_{l}^m$. } 
\begin{tabular}{lrrrrrrrrr} 
\hline 
 \hline 
 \\[-0.3cm] 
 ALFs  & $\Theta_{2}^{2,1}$ & $\Theta_{4}^{2,1}$ & $\Theta_{6}^{2,1}$ & $\Theta_{8}^{2,1}$ & $\Theta_{10}^{2,1}$ & $\Theta_{12}^{2,1}$ & $\Theta_{14}^{2,1}$ & $\Theta_{16}^{2,1}$ & $\Theta_{18}^{2,1}$ \\ 
\\[-0.3cm] 
 \hline 
 \\[-0.3cm] 
 $P_{2}^{2}$ & $ 0.969613$ & $-0.215005$ & $ 0.092839$ & $-0.051492$ & $ 0.032672$ & $-0.022555$ & $ 0.016500$ & $-0.012591$ & $ 0.009922$ \\ 
$P_{4}^{2}$ & $-0.243433$ & $-0.801859$ & $ 0.420597$ & $-0.248102$ & $ 0.161830$ & $-0.113384$ & $ 0.083680$ & $-0.064221$ & $ 0.050808$ \\ 
$P_{6}^{2}$ & $ 0.024257$ & $ 0.540019$ & $ 0.473174$ & $-0.437451$ & $ 0.333550$ & $-0.252143$ & $ 0.194322$ & $-0.153246$ & $ 0.123471$ \\ 
$P_{8}^{2}$ & $-0.001262$ & $-0.137140$ & $-0.696761$ & $-0.090980$ & $ 0.290658$ & $-0.306960$ & $ 0.276429$ & $-0.238222$ & $ 0.203051$ \\ 
$P_{10}^{2}$ & $ 0.000040$ & $ 0.019107$ & $ 0.314564$ & $ 0.673921$ & $-0.228327$ & $-0.054861$ & $ 0.176162$ & $-0.216336$ & $ 0.220315$ \\ 
$P_{12}^{2}$ & $-0.000001$ & $-0.001705$ & $-0.077477$ & $-0.495166$ & $-0.482704$ & $ 0.397208$ & $-0.172742$ & $ 0.009098$ & $ 0.087208$ \\ 
$P_{14}^{2}$ & $ 0.000000$ & $ 0.000107$ & $ 0.012402$ & $ 0.186257$ & $ 0.607874$ & $ 0.188568$ & $-0.384773$ & $ 0.307862$ & $-0.178620$ \\ 
$P_{16}^{2}$ & $-0.000000$ & $-0.000005$ & $-0.001412$ & $-0.045175$ & $-0.330239$ & $-0.600012$ & $ 0.111151$ & $ 0.223715$ & $-0.308380$ \\ 
$P_{18}^{2}$ & $ 0.000000$ & $ 0.000000$ & $ 0.000121$ & $ 0.007822$ & $ 0.112027$ & $ 0.470489$ & $ 0.458997$ & $-0.321769$ & $ 0.005637$ \\ 
$P_{20}^{2}$ & $-0.000000$ & $-0.000000$ & $-0.000008$ & $-0.001026$ & $-0.026937$ & $-0.214914$ & $-0.557134$ & $-0.219246$ & $ 0.384333$ \\ 
$P_{22}^{2}$ & $ 0.000000$ & $ 0.000000$ & $ 0.000000$ & $ 0.000106$ & $ 0.004909$ & $ 0.068230$ & $ 0.339372$ & $ 0.548788$ & $-0.048767$ \\ 
$P_{24}^{2}$ & $-0.000000$ & $-0.000000$ & $-0.000000$ & $-0.000009$ & $-0.000708$ & $-0.016319$ & $-0.138415$ & $-0.454789$ & $-0.430826$ \\ 
$P_{26}^{2}$ & $ 0.000000$ & $ 0.000000$ & $ 0.000000$ & $ 0.000001$ & $ 0.000083$ & $ 0.003084$ & $ 0.041965$ & $ 0.235739$ & $ 0.522540$ \\ 
$P_{28}^{2}$ & $-0.000000$ & $-0.000000$ & $-0.000000$ & $-0.000000$ & $-0.000008$ & $-0.000475$ & $-0.010002$ & $-0.088752$ & $-0.346057$ \\ 
$P_{30}^{2}$ & $ 0.000000$ & $ 0.000000$ & $ 0.000000$ & $ 0.000000$ & $ 0.000001$ & $ 0.000061$ & $ 0.001942$ & $ 0.026012$ & $ 0.160174$ \\ 
$P_{32}^{2}$ & $-0.000000$ & $-0.000000$ & $-0.000000$ & $-0.000000$ & $-0.000000$ & $-0.000007$ & $-0.000315$ & $-0.006184$ & $-0.056812$ \\ 
$P_{34}^{2}$ & $ 0.000000$ & $ 0.000000$ & $ 0.000000$ & $ 0.000000$ & $ 0.000000$ & $ 0.000001$ & $ 0.000043$ & $ 0.001226$ & $ 0.016225$ \\ 
$P_{36}^{2}$ & $-0.000000$ & $-0.000000$ & $-0.000000$ & $-0.000000$ & $-0.000000$ & $-0.000000$ & $-0.000005$ & $-0.000207$ & $-0.003851$ \\ 
$P_{38}^{2}$ & $ 0.000000$ & $ 0.000000$ & $ 0.000000$ & $ 0.000000$ & $ 0.000000$ & $ 0.000000$ & $ 0.000001$ & $ 0.000030$ & $ 0.000776$ \\ 
$P_{40}^{2}$ & $-0.000000$ & $-0.000000$ & $-0.000000$ & $-0.000000$ & $-0.000000$ & $-0.000000$ & $-0.000000$ & $-0.000004$ & $-0.000135$ \\ 
$P_{42}^{2}$ & $ 0.000000$ & $ 0.000000$ & $ 0.000000$ & $ 0.000000$ & $ 0.000000$ & $ 0.000000$ & $ 0.000000$ & $ 0.000000$ & $ 0.000021$ \\ 
$P_{44}^{2}$ & $-0.000000$ & $-0.000000$ & $-0.000000$ & $-0.000000$ & $-0.000000$ & $-0.000000$ & $-0.000000$ & $-0.000000$ & $-0.000003$ \\ 
$P_{46}^{2}$ & $ 0.000000$ & $ 0.000000$ & $ 0.000000$ & $ 0.000000$ & $ 0.000000$ & $ 0.000000$ & $ 0.000000$ & $ 0.000000$ & $ 0.000000$ \\
 \\ 
$\Lambda_n^{2,1}$ & $  11.1290$ & $  41.3329$ & $  91.0602$ & $ 160.4244$ & $ 249.4685$ & $ 358.2134$ & $ 486.6705$ & $ 634.8468$ & $ 802.7469$ \\ 
 \hline 
\end{tabular} 
\end{small}
\tablefoot{Symmetric modes with wave number $m = 2$ and spin parameter $\nu = 1$. Also shown are associated eigenvalues $\Lambda_n^{m,\nu}$.}
\end{table*}

In general, Laplace's tidal equation (\eq{equation_laplace}) cannot be solved analytically and must therefore be treated numerically. In practice, this can be done very efficiently using spectral methods \citep[e.g.][]{LC1969,Wang2016}. In the present work, the Hough functions are expanded as series of ALFs,
\begin{align}
& \HoughF{\nn}{\mm}{\spinpar}  = \sum_{\pp=0}^{N} \scal{\LegFnorm{\abs{\mm}+2\pp}{\mm} }{ \HoughF{\nn}{\mm}{\spinpar}} \LegFnorm{\abs{\mm}+2\pp}{\mm}  & \mbox{for} \ \nn-\abs{\mm} \ \mbox{even}, \\
& \HoughF{\nn}{\mm}{\spinpar} = \sum_{\pp=0}^{N} \scal{\LegFnorm{\abs{\mm}+2\pp+1}{\mm} }{ \HoughF{\nn}{\mm}{\spinpar}} \LegFnorm{\abs{\mm}+2\pp+1}{\mm}  & \mbox{for} \ \nn-\abs{\mm} \ \mbox{odd},
\end{align}
where $N$ denote the truncation degree of the series and $\pp$ is an integer. For a given order $\mm$ and spin parameter $\spinpar$, the expansion coefficients defining the Hough functions, $\HoughF{\nn}{\mm}{\spinpar}$, and their associated eigenvalues, $\Houghval{\nn}{\mm}{\spinpar}$, are computed simultaneously for all degrees $\nn$. The auxiliary functions $\Houghcol{\nn}{\mm}{\spinpar}$ and $\Houghlon{\nn}{\mm}{\spinpar}$, which describe the horizontal structure of the tidal flow, are then straightforwardly obtained from the set $\left\{ \HoughF{\nn}{\mm}{\spinpar} \right\}_{\nn}$ using \eqsthree{lapcol}{laplon}{hough_velocity}.


It is worth noting that the spin parameter, $\spinpar$, becomes independent of the tidal frequency in the asymptotic regime of rapid rotation, as illustrated by the Earth-Moon system. Since the Earth's spin angular velocity is much greater than the Moon's anomalistic mean motion, the semidiurnal tidal frequency satisfies $\ftide \approx 2 \spinrate$, implying $\spinpar \approx 1$. Hence, for the Earth's semidiurnal tide, $\spinpar$ can reasonably be assumed constant and set to unity throughout the entire evolution of the Earth-Moon system since its formation. As a consequence, the corresponding Hough functions need to be evaluated only once, despite large variations in the tidal frequency. 


For illustration, \fig{fig:hough_functions} displays the first Hough functions that are symmetric with respect to the equator ($\nn= 2 , 4 , 6, \ldots$), together with their associated velocity-field functions, for $\mm=2$ and $\spinpar=1$. Computations were performed with the \texttt{TRIP} dedicated computed algebra system \citep[][]{GL2011}. The Hough functions plotted in \fig{fig:hough_functions} are analogous to ALFs but distorted by planetary rotation. Table~\ref{tab:table_coefs} lists their series expansion coefficients and eigenvalues. Small discrepancies are observed between Table~\ref{tab:table_coefs} and Table~3.1 of \cite{LC1969}, beyond the third decimal place of the computed values. For example, we find  $\scal{\HoughF{2}{2}{1}}{ \LegFnorm{2}{2}} = 0.969613 $ (Table~\ref{tab:table_coefs}) instead of $\scal{\HoughF{2}{2}{1}}{ \LegFnorm{2}{2}} = 0.969152 $ \citep[][Table~3.1]{LC1969}. These differences may arise from numerical precision and from the truncation degree, $N$, used in the calculations. The values listed in Table~\ref{tab:table_coefs} were computed with double precision and $N=200$, whereas \citet{LC1969} likely solved Laplace's tidal equation using single precision and a smaller truncation degree, although this is not stated explicitly in the paper. 


\section{Calculation of the resonance frequency}
\label{app:calculation_resonance_freq}

We detail here the derivations leading to \eq{condition_resonance_ftide}. This condition is obtained by solving $D^2 \left( \ftide \right) = 0$, where $D^2 $ denotes the squared denominator of the integration constant parametrising the solution (\eq{Aconst_solution}), expressed as 
\begin{equation}
\label{D2_general}
D^2 \left( \ftide \right) = \abs{\kvert}^2 + \left( \betasurf \Lambn - \frac{1}{2} \right)^2 - 2 \left( \betasurf \Lambn - \frac{1}{2} \right) \Im \left( \kvert \right), 
\end{equation}
and $\kvert^2$ the squared vertical wavenumber, given by 
\begin{equation}
\label{kvert2_general}
\kvert^2 = \gammapar \betasurf \Lambn - \frac{1}{4}. 
\end{equation}
In the above equation, $\kvert^2$ can be either positive or negative. Therefore, we should treat the two cases separately. 

\subsection{Case $\kvert^2 >0$}

For $\kvert^2 >0$, the squared modulus of the vertical wavenumber and its imaginary component simplifies to 
\begin{align}
& \abs{\kvert}^2 = \kvert^2, & \Im \left\{ \kvert \right\} =0,
\end{align}
thus allowing \eq{D2_general} to be rewritten as
\begin{equation}
\label{D2_case_pos}
D^2 \left( \ftide \right) = \kvert^2 + \left( \betasurf \Lambn -\frac{1}{2} \right)^2.
\end{equation}
Substituting $\kvert^2$ with \eq{kvert2_general} in \eq{D2_case_pos}, we end up with
\begin{equation}
D^2 \left( \ftide \right) = \betasurf \Lambn \left( \betasurf \Lambn + \gammapar - 1  \right).
\end{equation}
As a consequence, the condition $D^2 \left( \ftide \right) = 0 $ implies that 
\begin{equation}
\betasurf = \frac{1 - \gammapar}{\Lambn},
\end{equation}
which, using the definition of $\betasurf$ (\eq{dimensionless_numbers}), $\betasurf = \ggravi \heightsurf / \left( \ftide^2 \Rpla^2 \right)$, yields the relationship given by \eq{condition_resonance_ftide},
\begin{equation}
\label{condition_resonance_ftide_positive}
\ftide^2 = \frac{\ggravi \heightsurf \Lambn}{\Rpla^2 \left( 1 - \gammapar \right) }.
\end{equation}

\subsection{Case $\kvert^2 <0$}
For $\kvert^2<0$, the squared modulus and imaginary component of the vertical wavenumber are given by 
\begin{align}
& \abs{\kvert}^2 = - \kvert^2, & \Im \left\{ \kvert \right\} = \sqrt{- \kvert^2}.
\end{align}
Substituting these expressions in \eq{D2_general}, we obtain
\begin{equation}
\label{D2_case_negative}
D^2 \left( \ftide \right)  = \left( \sqrt{\frac{1}{4} - \gammapar \betasurf \Lambn} - \betasurf \Lambn + \frac{1}{2} \right)^2 .
\end{equation}
We note that solutions of the equation $D^2 \left( \ftide \right) = 0$ have to satisfy the condition
\begin{equation}
\label{cond_case_neg}
\betasurf \Lambn \geq \frac{1}{2}.
\end{equation}
Assuming that it is the case, we rewrite \eq{D2_case_negative} as
\begin{equation}
\frac{1}{4} - \gammapar \betasurf \Lambn  = \left( \betasurf \Lambn - \frac{1}{2} \right)^2 ,
\end{equation}
which eventually simplifies to 
\begin{equation}
\betasurf \Lambn \left( \betasurf \Lambn + \gammapar- 1 \right) = 0,
\end{equation}
and we retrieve \eq{condition_resonance_ftide_positive}. It is noteworthy that the condition given by \eq{cond_case_neg} is satisfied for $\Lambn>0$. In the isentropic configuration, $\gammapar =0$, meaning that $\betasurf \Lambn = 1$. In the isothermal configuration, $\gammapar = \rcp $ with $\rcp = \left( \adiabexp - 1 \right)/\adiabexp$, which leads to $\betasurf \Lambn = 1/\adiabexp$. The maximal theoretical value of $\adiabexp$ is attained for monoatomic gases: $\adiabexp=5/3 < 2$ \citep[e.g.][Eqs.~(44.1) and~(44.2), p.~124, with $l=3$]{LL1969book}. If $\Lambn <0$, the equation $D^2 \left( \ftide \right) =0$ has no real solution, which means that the mode cannot be resonantly excited by the tidal forcing.

\end{document}